\documentclass[%
 reprint,
showpacs,
aps,prl]{revtex4-1}

\usepackage{graphicx}
\usepackage{dcolumn}
\usepackage{bm}

\begin{document}
\title{\textbf{\centering First-principles prediction of phonon-mediated superconductivity in $X$BC ($X$= Mg, Ca, Sr, Ba)}}

\author{Enamul Haque}
\author{M. Anwar Hossain}%
\email{anwar647@mbstu.ac.bd, enamul.phy15@yahoo.com}
\affiliation{Department of Physics, Mawlana Bhashani Science and Technology University 
	\\ Santosh, Tangail-1902, Bangladesh}

\author{Catherine Stampfl}
\email{catherine.stampfl@sydney.edu.au}
\affiliation{School of Physics, The University of Sydney, Sydney,
	\\ New South Wales, 2006, Australia.
}
\date{\today}
\begin{abstract}
From first-principles calculations, we predict four new intercalated hexagonal $X$BC ($X$=Mg, Ca, Sr, Ba) compounds to be dynamically stable and phonon-mediated superconductors. These compounds form a LiBC like structure but are metallic.  The calculated superconducting critical temperature, $T{_c}$, of MgBC is 51 K. The strong attractive interaction between $\sigma$-bonding electrons and the B${_{1g}}$ phonon mode  gives rise to a larger electron-phonon coupling constant (1.135) and hence high $T_c$; notably, higher than that of MgB$_2$. The other compounds have a low superconducting critical temperature (4-17 K)  due to the interaction between $\sigma$-bonding electrons and low energy phonons (E${_{2u}}$ modes). Due to their energetic and dynamic stability, we envisage that these compounds can be synthesized experimentally.  
\end{abstract}


\maketitle

Hexagonal layered MgB$_2$ is a well-known phonon-mediated superconductor with a $T_c=39$ K \cite{nagamatsu2001superconductivity}. In MgB$_2$, the $\sigma$-band crosses the Fermi level and hybridization with other conduction electrons is weak. The high $T_c$-state of this material develops from the strong attractive interaction between the electrons of the $\sigma$-band and the E$_{2g}$ mode of vibrations. Likely MgB$_2$, materials intercalated with alkali (earth) metals show superconductivity with $T_c$ much smaller than that of MgB$_2$ \cite{hannay1965superconductivity,belash1989superconductivity,emery2005superconductivity,calandra2005theoretical,csanyi2005role,profeta2012phonon,sanna2007anisotropic,gauzzi2007enhancement,tiwari2017superconductivity,nishiyama2017preparation,liao2017phonon,Tawfik}.  Many studies have proposed possible ways to improve the transition temperature through carbon (or others) doping \cite{bharathi2003superconductivity,jemima2003effect,balaselvi2003systematics,kazakov2005carbon}. However, the $T_c$ has not been  found  to improve significantly  \cite{bharathi2003superconductivity,jemima2003effect,balaselvi2004stoichiometric,kazakov2005carbon,ohmichi2004enhancement,bharathi2002carbon}. The $T_c$ decreases on carbon substitution of B in MgB$_2$ due to the introduction simultaneous disorder by carbon \cite{kazakov2005carbon, braccini2005high,pissas2004vortex,ohmichi2004enhancement}. Thus, it may be an alternative to examine the superconducting properties by synthesizing pure MgBC compounds. 
\newline
\newline 
From the first-principles investigations of LiBC with hole-doping, it was found to exhibit superconductivity below 100 K \cite{rosner2002prediction}. Unfortunately, experimentalists  have not identified superconductivity in it due to induced structural distortions \cite{fogg2003libc,fogg2006chemical,souptel2003synthesis,bharathi2002synthesis,gao2015prediction}. This is a similar effect to that of carbon doping in MgB$_2$. Recently, LiB$_{1+x}$C$_{1-x}$ materials have been predicted to be superconductors, like MgB$_2$ \cite{Qielectron}. Also on the basis of first-principles calculations, Gao $et$ $al.$ reported Li$_3$B$_4$C$_2$ (also Li$_2$B$_3$C) to be MgB$_2$ like superconductor with $T_c \sim 53$ K \cite{gao2015prediction}.  Since then, another phase, Li$_4$B$_5$C$_3$, has been reported from ab initio studies to be a superconductor with a transition temperature of 16.8 K  \cite{bazhirov2014electron}. 
\begin{figure}[b]
	\includegraphics[width=4cm, height=5cm]{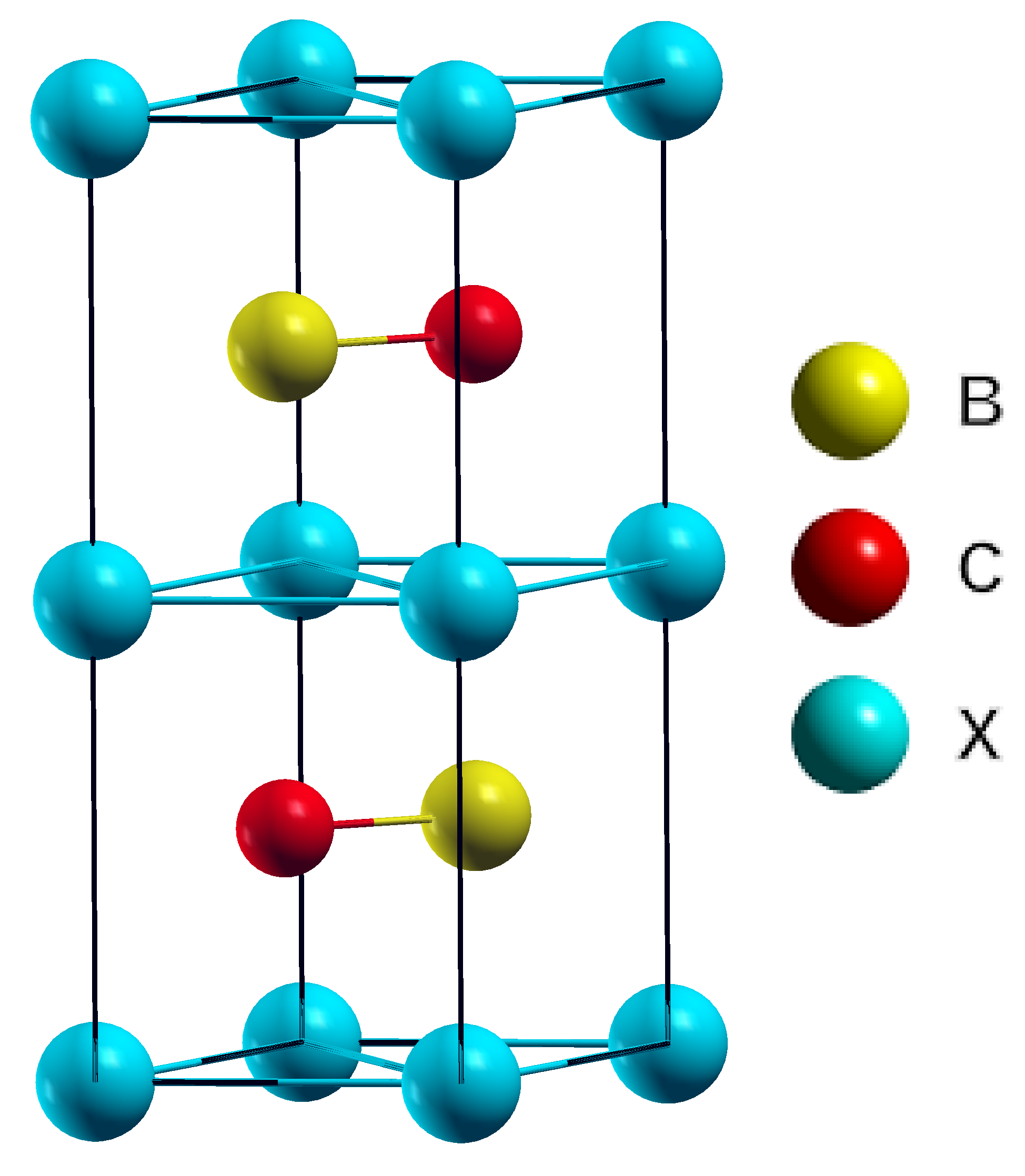}
	\caption{\label{fig:} The ground state crystal structure of $X$BC ($X$=Mg, Ca, Sr, Ba). The $X$, B, and C atoms are indicated by cyan, yellow, and red spheres, respectively.}
\end{figure}
From the first-principles study of NaB$_{1+x}$C$_{1-x}$, Miao $et$ $al.$ predicted that it would be more promising superconductor than LiB$_{1-x}$C$_x$ \cite{miao2016first}. Like LiBC, Ravindran $et$ $al.$ predicted that hole-doped MgB$_2$C$_2$ is a potential superconducting material \cite{ravindran2001detailed}. Since high-quality single crystals of LiBC have already been  synthesized  [16,17], it may be possible to synthesize $X$BC ($X$=Mg, Ca, Sr, Ba).  
\begin{figure*}
	\includegraphics[trim={0.83cm 0 0 0},clip, width=18.2cm, height=15.2cm]{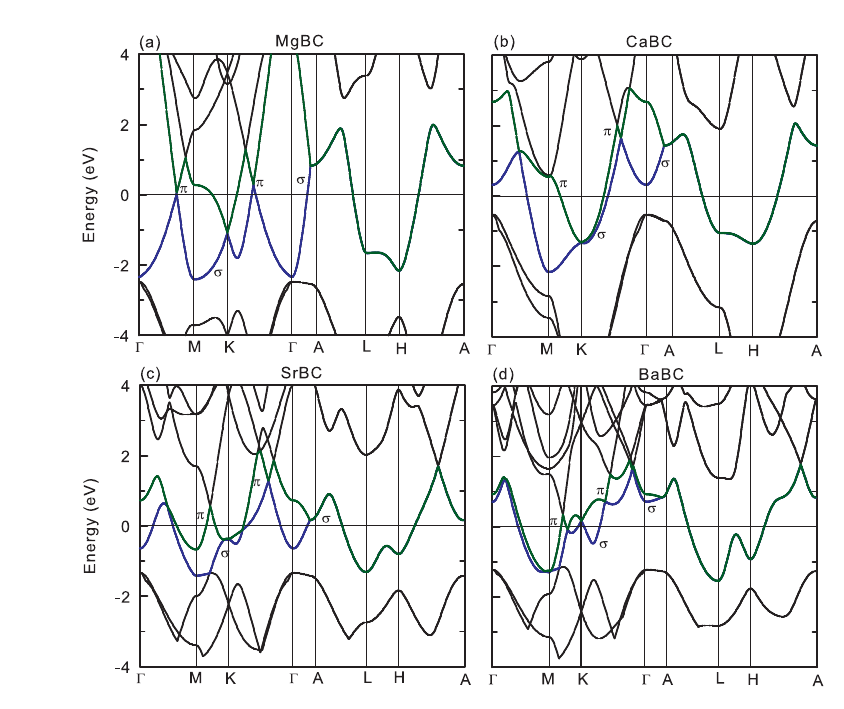}
	\caption{\label{fig:} Electronic band structure of the four compounds: (a) MgBC,  (b) CaBC, (c) SrBC, and (d) BaBC. The Fermi level is the zero of energy. The blue and red curves with symbols represent the energy bands that cross the Fermi level and form the Fermi surface. We have selected high-symmetry \textbf{k}-points in the Brillouin zone and the values of them in fractional coordinates are $\Gamma (0,0,0)$, M (1/2,0,0), K (1/3,1/3,0), A (0,0,1/2), L (1/2,0,1/2), and H (1/3,1/3,1/2).  }
\end{figure*}
In the present paper, we report, using the first-principles calculations, four new superconducting stoichiometric compounds ($X$BC ($X$=Mg, Ca, Sr, Ba)) that are dynamically stable and may be synthesized from constituent elemental solids. We find that strong electron-phonon interactions exist in all these materials. The MgBC structure has a predicted  $T_c (\sim 51$ K) higher than that of MgB$_2$; while SrBC and BaBC have a superconducting state below $\sim$ 17 K (The CaBC structure has very small $T_c ( 4$ K)).  
\newline
All calculations were performed using the plane wave pseudopotential approach and the generalized gradient approximation of Perdew-Burke-Ernzerhof (PBE-GGA) \cite{perdew1996generalized,perdew2008restoring}  for the exchange correlation functional, as implemented in Quantum Espresso  \cite{giannozzi2009quantum}. We use the ultrasoft pseudopotentials of Vanderbilt  \cite{vanderbilt1990soft} and perform full structural relaxation. After optimizing the \textbf{k}-point mesh and cutoff energy, we selected a $12\times12\times4$ \textbf{k}-point mesh for self-consistent field calculations, a 50 Ry cutoff energy for the wave functions, and a 400 Ry energy cutoff for the charge density. For the phonon calculations, we use a 662 grid of uniform \textbf{q}-points and the same \textbf{k}-point mesh as above \cite{textMgB2}. We used a finer  $18\times18\times6$ \textbf{k}-point mesh for the calculation of the electron-phonon (e-ph) linewidth and e-ph coupling constants employing the optimized tetrahedron method in e-ph calculation  \cite{kawamura2014improved}.  We performed the phonon calculations using density functional perturbation theory (DFPT) of linear response \cite{baroni2001phonons}.  For the electron-phonon coupling constant (EPC) calculations, we used the Migdal-Eliashberg formulism  \cite{eliashberg1960interactions}.
\begin{figure}[t]
	\includegraphics[width=8.6cm, height=9cm]{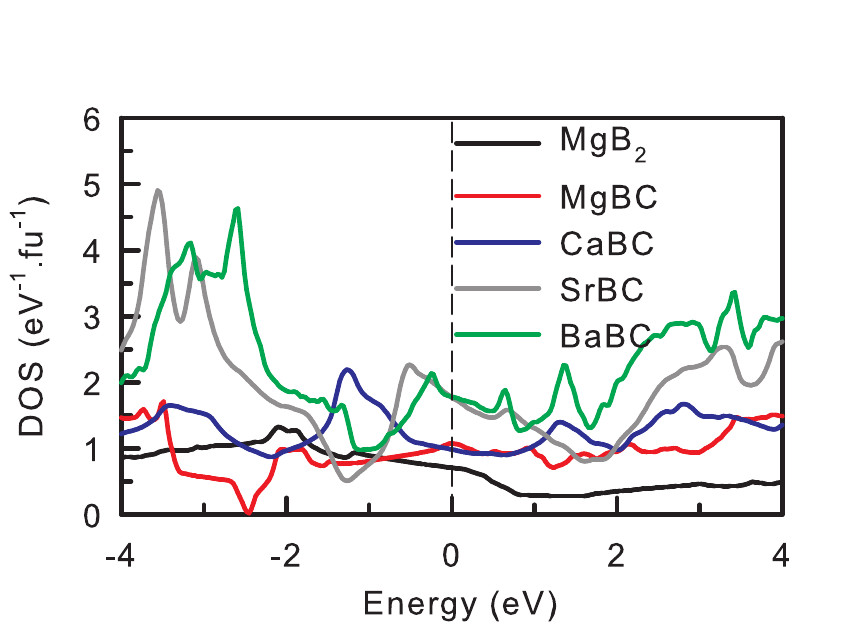}
	\caption{\label{fig:} Calculated total density of states (DOS) of $X$BC ($X$=Mg, Ca, Sr, Ba) within the energy range from $-$4 to 4. The Fermi level is the zero of energy and indicated by the vertical dashed line. }
\end{figure}
 In this formulism, the Eliashberg spectral function is defined as \cite{allen1972neutron,allen1975transition}
\begin{equation}
\alpha^2 F(\omega)=\frac{1}{2\pi N(E_F)} \sum_{\boldsymbol{qv}} \delta(\omega-\omega_{\boldsymbol{qv}})\frac{\gamma_{qv}}{\hbar\omega_{\boldsymbol{qv}}}
\end{equation}
where $N(E_F)$ is the density of states at the Fermi level and $\gamma_{\boldsymbol{qv}}$ is the electron-phonon linewidth for wave vectors $\boldsymbol{q}$ and $\boldsymbol{v}$.  The EPC is determined  by  \cite{allen1972neutron,allen1975transition}
\begin{equation}
λ=2\int \frac{\alpha^2 F(\omega)}{\omega} d\omega .
\end{equation}

Using the calculated EPC, the superconducting transition temperature is evaluated by the Allen-Dynes equation \cite{allen1972neutron,allen1975transition}
\begin{equation}
T_c=\frac{\omega_{_{ln}}}{1.2} \exp \left[\frac{(-1.04(1+\lambda))}{(\lambda(1-0.62\mu^* )-\mu^*)}\right]
\end{equation} 

where $\mu^*$ stands for the Coulomb pseudopotential constant and its value ranges between 0.1 and 0.15  \cite{richardson1997high,lee1995first}.  $\omega_{ln}$ stands for the logarithmic average frequency and is defined as \cite{allen1972neutron,allen1975transition}
\begin{equation}
\omega_{ln}=\exp \left[\frac{2}{\lambda} \int \frac {d\omega}{\omega} \alpha^2 F(\omega)  ln(\omega) \right].
\end{equation}

The crystal structure of $X$BC ($X$=Mg, Ca, Sr, Ba) is similar to that of MgB$_2$. $X$ atoms have no bonds with either B or C. Boron and carbon are bonded together in a primitive fashion. Unlike MgB$_2$, the unit cell contains six equivalent atoms, two of each species. Figure 1 shows the hexagonal crystal structure of  $X$BC ($X$=Mg, Ca, Sr, Ba).
 \begin{table}[t]
 	\caption{\label{tab:table1}%
 		Calculated fully relaxed lattice parameters of MgB$_2$ and $X$BC of ($X$=Mg, Ca, Sr, Ba).
 	}
 	\begin{ruledtabular}
 		\begin{tabular}{lcdr}
 			\textrm{Compounds}&
 			\textrm{$a$(\AA)}&
 			\textrm{\centering$c/a$}&
 			\\
 			\colrule
 			MgB$_2$ & 3.081 & 1.145 &\\
 			MgBC & 2.808 & 2.608 &\\
 			CaBC & 2.951 & 2.745 &\\
 			SrBC & 3.013 & 2.973 &\\
 			BaBC & 3.084 & 3.190 &\\
 		\end{tabular}
 	\end{ruledtabular}
 \end{table}
Our theoretical value of the lattice parameters of MgB$_2$ is in a good agreement with the experimental value ($a=3.086 \AA$ and $c/a=1.42)$ \cite{nagamatsu2001superconductivity}, as listed in table 1. The fully relaxed lattice constant along the $a$-axis is close to the value of MgB$_2$ while that along the $c$-axis of all compounds become around twice that of MgB$_2$ (see Table 1). We have found that all the studied compounds are energetically stable \cite{enamultext}. 
The electronic band structures of the four compounds are shown in figure 2. All the compounds possess metallic character. We see that two energy bands cross the Fermi level and form the Fermi surface for all materials. The symbols, $\sigma$ and $\pi$, indicate the type of bonding electrons belonging to the energy bands crossing the Fermi level. These bands are highly dispersive along all directions except L-H. These bands are slightly flat along A-L above $E_F$ and L-H below $E_F$. The bands crossing the Fermi level are doubly degenerate along the $\Gamma$-A direction. 
\newline  
If we compare the band structure of MgBC and CaBC, we see that the Fermi level in CaBC is shifted to higher energy than that of MgBC. Therefore, this may lead to a reduction of $T_c$ or eliminate the superconducting state of CaBC. However, the shift in energy of SrBC and BaBC is small as compared to CaBC. 
 
\begin{figure}[h]
	\includegraphics[trim={0 0.15cm 0 0},clip, width=8.6cm, height=8cm]{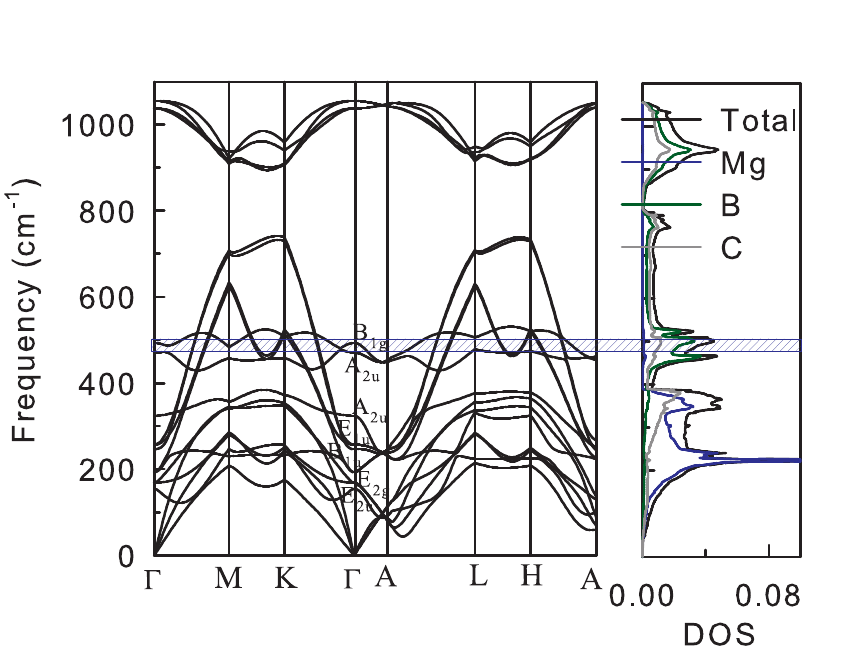}
	\includegraphics[trim={0 0.15cm 0 0.7cm},clip, width=8.6cm, height=7.3cm]{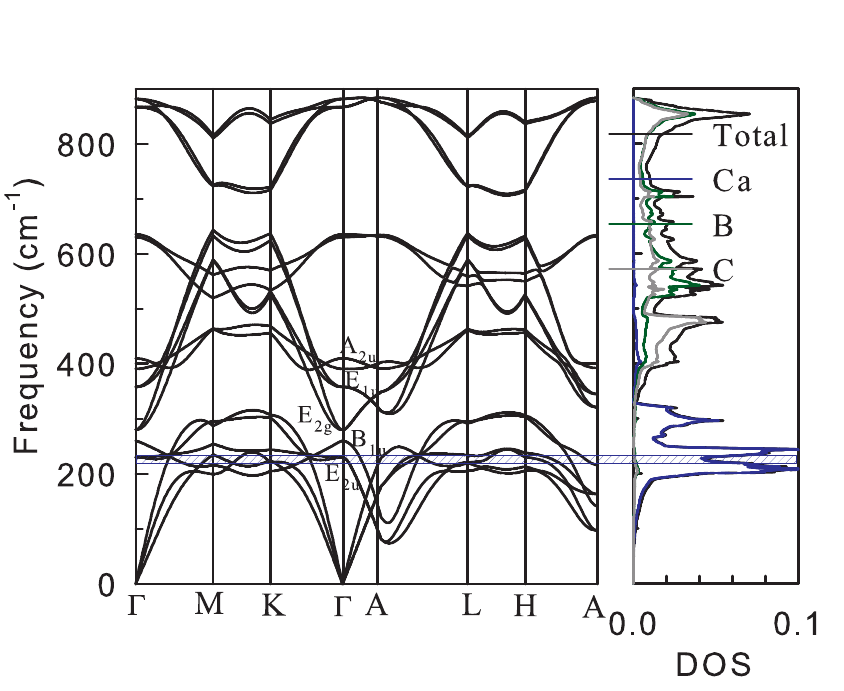}
	\caption{\label{fig:} Phonon dispersion relations, total, and atom projected phonon density of states of MgBC (top panel) and CaBC (bottom panel). The blue shaded area indicates the modes of the region where electrons are strongly coupled.}
\end{figure}
Figure 3 shows the calculated total density of states of $X$BC ($X$=Mg, Ca, Sr, Ba). The density of states at the Fermi level has increased significantly in all the studied compounds compared to MgB$_2$. In the case of CaBC, the DOS at the Fermi level is similar to MgBC and notably lower than both of SrBC and BaBC. As mentioned above, the DOS of MgBC at the Fermi level is much increased in comparison with that of MgB$_2$ for the electron injected by carbon \cite{mazin2003electronic}. Since the $\sigma$-band crosses the Fermi level, $\sigma$-bonding electrons can strongly attract  certain modes of vibration, like the E$_{2g}$ mode of MgB$_2$. Therefore, all compounds should be superconductors at a certain temperature. We will now investigate the electron-phonon coupling within Migdal-Eliashberg formulism. 
\\
\\
\begin{figure}[h]
	\includegraphics[trim={0 0.15cm 0 0},clip, width=8.6cm, height=8cm]{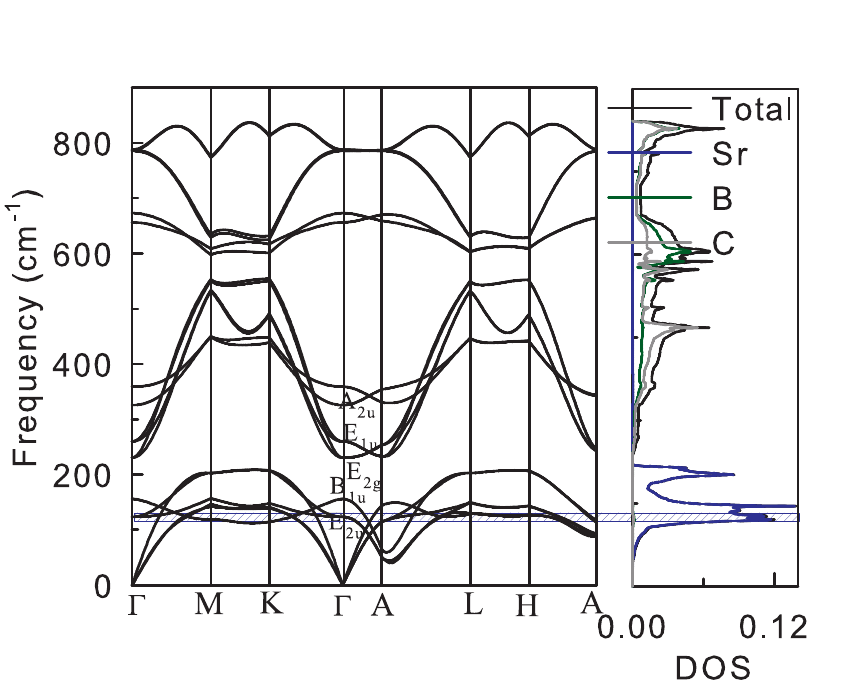}
	\includegraphics[trim={0 0.15cm 0 0.7cm},clip, width=8.6cm, height=7.3cm]{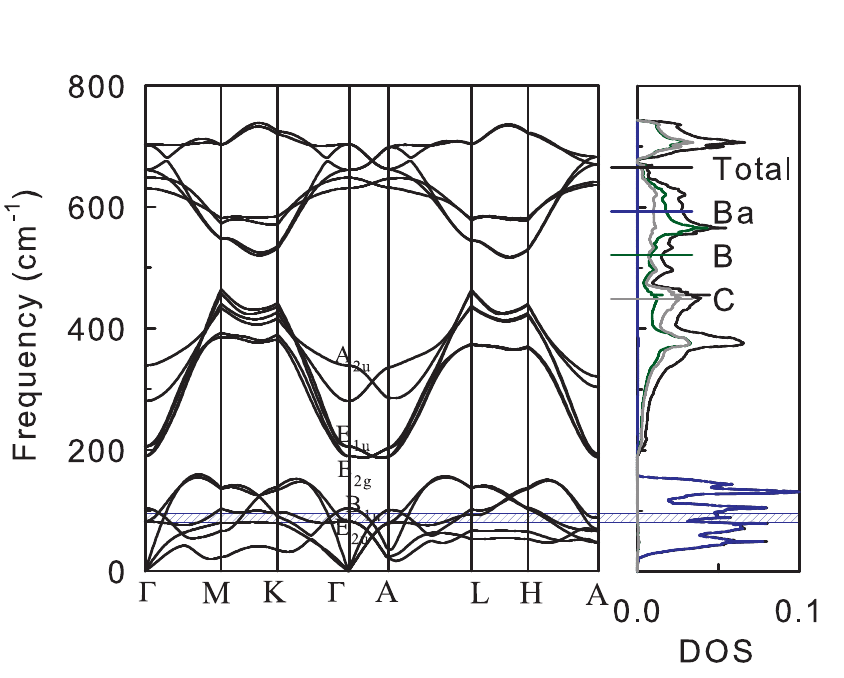}
	\caption{\label{fig:} Phonon band structures, total, and atom projected phonon density of states of SrBC (top panel) and BaBC (bottom panel). The blue shaded area indicates the modes of the region (E$_{2u}$) where electrons are strongly coupled. }
\end{figure} 
The dynamical stability of a crystal is an important criterion in order to be able to synthesize it. Many researchers have reported the failure of synthesis of potentially promising materials due to dynamical instability  \cite{grimvall2012lattice}. The phonons determine the dynamical stability of a crystal; if any imaginary frequencies appears in the phonon band structure, the crystal structure is dynamically unstable. Recently, Kato $et$ $al$. reported the possibility of MgB$_2$-like superconductivity in MgB$Y$ ($Y$=C, Be, Li) from electronic band structure analysis without considering the dynamical stability of these compounds  \cite{kato2004possibility}. They also predicted that MgBC might not be a two-band superconductor because the $\sigma$-band is completely filled. We note that the lattice constant of dynamically stable MgBC along the $c$-axis is twice as large as that of considered the one in their study \cite{kato2004possibility}, i.e., it has different atomic structure. The Fermi level in MgBC is raised to higher energy, even around the sigma anti-bonding states, compared to that of MgB$_2$. In this respect, MgBC is different in electronic structure to that of MgB$_2$. 
\newline
\newline
Figure 4 shows the phonon dispersion relation and phonon density of states (DOS) of MgBC (upper panel) and CaBC (lower panel). We see that no imaginary frequency appears in the phonon band structure for both compounds. Therefore, hexagonal MgBC and CaBC are dynamically stable \cite{latticesta}. Low energy phonons mainly arise from Mg and Ca in MgBC, and CaBC respectively. Higher energy phonons arise from B and C in both compounds. In MgBC, the optical $\Gamma$-center modes are shifted to higher frequencies as compared to MgB$_2$. In contrast to MgB$_2$ (where the in-plane boron mode is E$_{2g}$), the B-C in-plane phonon at the $\Gamma$-point is the B$_{1g}$ mode  where electrons are strongly coupled, as indicated by the shaded area in figure 4. The shaded area corresponds to the highest peak region in the Eliashberg spectral function. The corresponding frequency of the B$_{1g}$ mode at the $\Gamma$-point is 495 cm$^{-1}$, which is much smaller than that of the value of 692 cm$^{-1}$ for MgB$_2$  \cite{kong2001electron}. 

For CaBC, low energy phonons of the E$_{2u}$ mode region arise from Ca atom and in this region, electrons are strongly coupled (maximum peak region in the Eliashberg spectral function). Higher energy phonons arising from B and C have very small contributions to electron-phonon interactions. Since the density of states (DOS) at the Fermi level of CaBC (see figure 3) is reduced as compared to MgBC , and the phonon frequency too (blue shaded region), $\sigma$-band electrons are not so strongly coupled with E$_{2u}$ modes of the phonons as like in MgB$_2$. Therefore, CaBC cannot be a MgB$_2$-like superconductor. 
\newline
\newline
The phonon energy of the blue shaded area at $\Gamma$-point is more reduced in SrBC and BaBC compared to CaBC, as shown in figure 5. Like CaBC, SrBC, and BaBC have E$_{2u}$ modes region where electrons are strongly coupled. However, unlike CaBC, the density of states of SrBC and BaBC at the Fermi level are notably higher.  Therefore, electrons should be more strongly coupled than those in CaBC. 
\begin{figure}[h]
	\centering
\includegraphics[trim={0 0.15cm 0 0.95cm},clip, width=8.6cm, height=4.62cm]{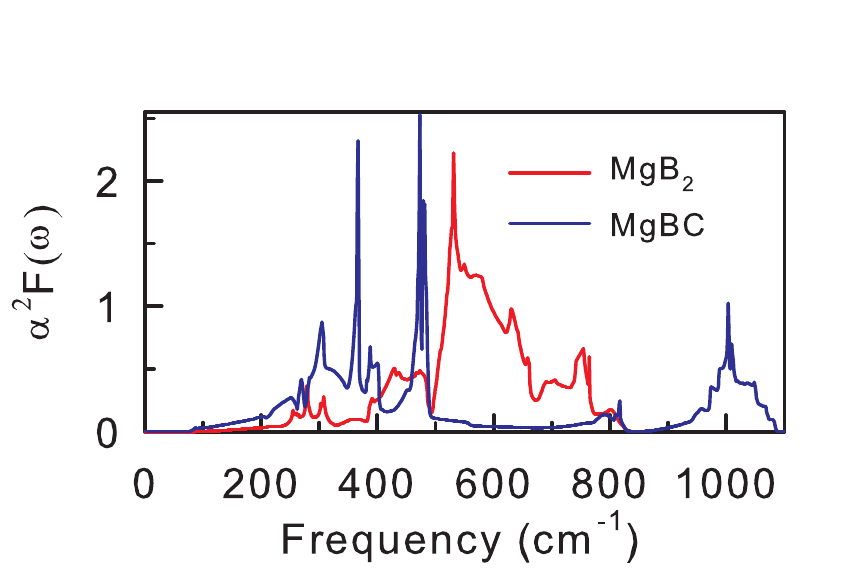}
\includegraphics[trim={0 0.15cm 0 0.75cm},clip, width=8.6cm, height=4.62cm]{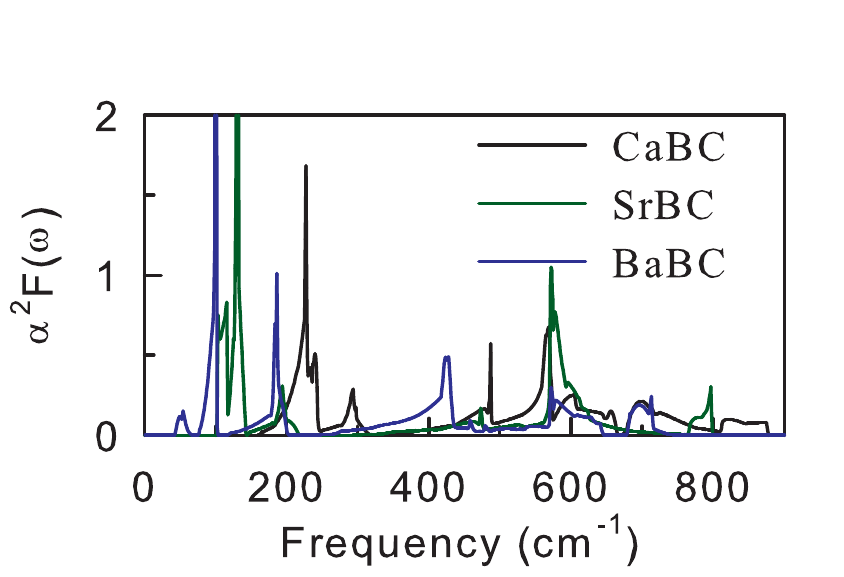}
\centering
\includegraphics[trim={0 0.15cm 0 0.7cm},clip, width=8.6cm, height=4cm]{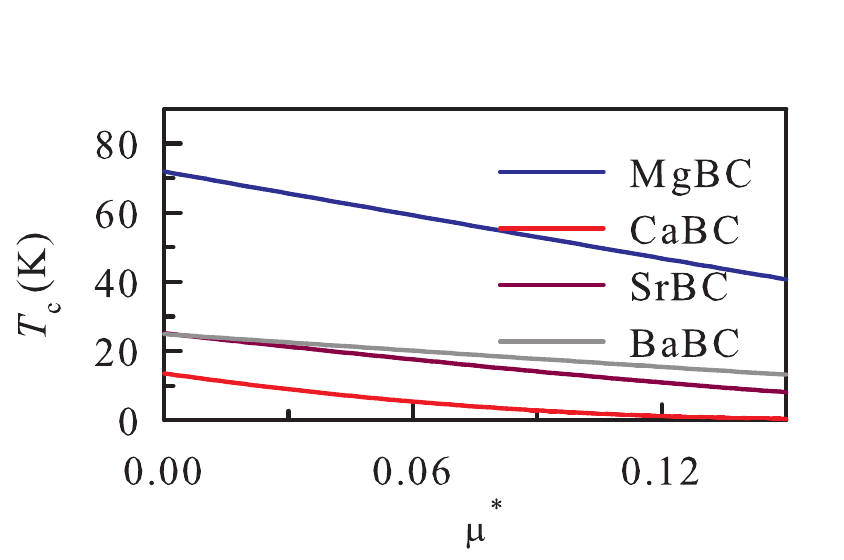}
	\caption{\label{fig:} Calculated Eliashberg spectral function $\alpha^2 F(\omega)$ of MgB$_2$ and MgBC (upper panel), and of CaBC, SrBC, BaBC (middle panel). The lower panel shows calculated superconducting transition temperature as a function of $\mu^*$.  }
\end{figure}
For hexagonal SrBC and BaBC, we do not obtain any imaginary frequencies. Therefore, both structures are energetically \cite{enamultext} and dynamically stable \cite{latticesta}. 
The phonon density of states of MgBC exhibits three distinct peaks; in contrast MgB$_2$ has only one peak  [18]. The first peak arises from the Mg of E$_{1u}$  mode and the second and third peaks arise from B and C. The second peak shows that B-has a dominant contribution. The Eliashberg spectral function of MgBC also shows three peaks, in comparison with just one peak of MgB$_2$.  The main peak around 694 cm$^{-1}$ arises from the predominant interaction between $\sigma$-band electrons and the B$_{1g}$ phonon mode. These three peaks in the Eliashberg spectral function are shifted to lower energy phonons (E$_{2u}$) in the case of the other remaining three compounds. For CaBC, the highest peak height is almost half of the highest peak height for MgBC, SrBC, and BaBC. From the Eliashberg spectral function, we can calculate the electron-phonon coupling constant, logarithmic average phonon frequency and hence the superconducting transition temperature using the Allen-Dynes equation  \cite{allen1975transition}. Our calculated superconducting parameters of the four compounds are listed in Table 2. In the table, we have used the value of the Coulomb pseudopotential to be 0.1.  

\begin{table}[t]
	\caption{\label{tab:table1}%
		Calculated superconducting parameters of fully relaxed structures $X$BC of ($X$=Mg, Ca, Sr, Ba). We have used the value of $µ^*$ to be 0.1 in Eqn. (3).
	}
	\begin{ruledtabular}
		\begin{tabular}{lcdr}
			\textrm{Compounds}&
			\textrm{$\omega_{ln}$(K)}&
			\textrm{$\lambda$}&
			\textrm{$T_c$(K)}\\
			\colrule
			MgBC & 610.05 & 1.135 & 51\\
			CaBC & 723.09 & 0.377 & 4\\
			SrBC & 382.27 & 0.693 & 13\\
			BaBC & 231.13 & 1.034 & 17\\
		\end{tabular}
	\end{ruledtabular}
\end{table}
For MgB$_2$, our calculated electron-phonon coupling constant ($\lambda$) is 0.844, in a good agreement with the available results \cite{kong2001electron,gao2015prediction} but slightly smaller than that of the value reported in Ref. \cite{yildirim2001giant}. This value is slightly larger than that of the obtained value by using Wannier interpolation method \cite{eiguren2008wannier,calandra2010adiabatic,margine2013anisotropic}. We find that the logarithmic average phonon frequency ($\omega_{ln}$) is 708 K, in a good agreement with available data \cite{gao2015prediction,an2001superconductivity,kong2001electron,choi2002first,choi2002origin,bohnen2001phonon,morshedloo2015first}. We obtain $T_c=37$K by using $\mu^*=0.1$, slightly smaller than that of the experimental value (39 K). We see that the maximum superconducting transition temperature is obtained for MgBC and minimum for CaBC. The electron-phonon coupling constant (1.135) of MgBC is a 26\% larger  than that obtained for MgB$_2$ (0.87-0.88)  \cite{gao2015prediction,kong2001electron,yildirim2001giant}. Thus, if we use the value of $\mu^*$ to be 0, we obtain $T_c$=72 K for MgBC. Even if we use the value of 0.15 for $\mu^*$, we still obtain a superconducting transition temperature above 40 K. Therefore, MgBC is a phonon-mediated superconductor with larger electron-phonon coupling constant and higher transition temperature than MgB$_2$.  The larger electron-phonon coupling constant of MgBC arises from three peaks in the phonon density of states, mainly due to the strong coupling of $\sigma$-band electrons with the B$_{1g}$ phonon mode. The others three compounds have a lower transition temperature compared to MgB$_2$.

In summary, we have predicted four new superconductors using the first-principles calculations. Hexagonal $X$BC ($X$=Mg, Ca, Sr, Ba) compounds are found to be phonon-mediated superconductors. Among these compounds, the calculated $T_c$ of MgBC is 51 K. The strong coupling between $\sigma$-bonding electrons and the B$_{1g}$ phonon mode gives rise to a larger electron-phonon coupling and hence high $T_c$. Thus, MgBC is a superconductor with $T_c$ higher than that of MgB$_2$. The other compounds have a low superconducting transition temperature due to the interaction between $\sigma$-bonding electrons and low energy phonons (E$_{2u}$ modes).

\begin{acknowledgments}
We must thank W.E Pickett for giving some helpful suggestions and critical reading of the manuscript.
\end{acknowledgments}

\nocite{*}

\bibliography{apssamp}

\providecommand{\noopsort}[1]{}\providecommand{\singleletter}[1]{#1}%
\begin{thebibliography}{60}%
\makeatletter
\providecommand \@ifxundefined [1]{%
 \@ifx{#1\undefined}
}%
\providecommand \@ifnum [1]{%
 \ifnum #1\expandafter \@firstoftwo
 \else \expandafter \@secondoftwo
 \fi
}%
\providecommand \@ifx [1]{%
 \ifx #1\expandafter \@firstoftwo
 \else \expandafter \@secondoftwo
 \fi
}%
\providecommand \natexlab [1]{#1}%
\providecommand \enquote  [1]{``#1''}%
\providecommand \bibnamefont  [1]{#1}%
\providecommand \bibfnamefont [1]{#1}%
\providecommand \citenamefont [1]{#1}%
\providecommand \href@noop [0]{\@secondoftwo}%
\providecommand \href [0]{\begingroup \@sanitize@url \@href}%
\providecommand \@href[1]{\@@startlink{#1}\@@href}%
\providecommand \@@href[1]{\endgroup#1\@@endlink}%
\providecommand \@sanitize@url [0]{\catcode `\\12\catcode `\$12\catcode
  `\&12\catcode `\#12\catcode `\^12\catcode `\_12\catcode `\%12\relax}%
\providecommand \@@startlink[1]{}%
\providecommand \@@endlink[0]{}%
\providecommand \url  [0]{\begingroup\@sanitize@url \@url }%
\providecommand \@url [1]{\endgroup\@href {#1}{\urlprefix }}%
\providecommand \urlprefix  [0]{URL }%
\providecommand \Eprint [0]{\href }%
\providecommand \doibase [0]{http://dx.doi.org/}%
\providecommand \selectlanguage [0]{\@gobble}%
\providecommand \bibinfo  [0]{\@secondoftwo}%
\providecommand \bibfield  [0]{\@secondoftwo}%
\providecommand \translation [1]{[#1]}%
\providecommand \BibitemOpen [0]{}%
\providecommand \bibitemStop [0]{}%
\providecommand \bibitemNoStop [0]{.\EOS\space}%
\providecommand \EOS [0]{\spacefactor3000\relax}%
\providecommand \BibitemShut  [1]{\csname bibitem#1\endcsname}%
\let\auto@bib@innerbib\@empty
\bibitem [{\citenamefont {Nagamatsu}\ \emph {et~al.}(2001)\citenamefont
  {Nagamatsu}, \citenamefont {Nakagawa}, \citenamefont {Muranaka},
  \citenamefont {Zenitani},\ and\ \citenamefont
  {Akimitsu}}]{nagamatsu2001superconductivity}%
  \BibitemOpen
  \bibfield  {author} {\bibinfo {author} {\bibfnamefont {J.}~\bibnamefont
  {Nagamatsu}}, \bibinfo {author} {\bibfnamefont {N.}~\bibnamefont {Nakagawa}},
  \bibinfo {author} {\bibfnamefont {T.}~\bibnamefont {Muranaka}}, \bibinfo
  {author} {\bibfnamefont {Y.}~\bibnamefont {Zenitani}}, \ and\ \bibinfo
  {author} {\bibfnamefont {J.}~\bibnamefont {Akimitsu}},\ }\href@noop {}
  {\bibfield  {journal} {\bibinfo  {journal} {nature}\ }\textbf {\bibinfo
  {volume} {410}},\ \bibinfo {pages} {63} (\bibinfo {year} {2001})}\BibitemShut
  {NoStop}%
\bibitem [{\citenamefont {Hannay}\ \emph {et~al.}(1965)\citenamefont {Hannay},
  \citenamefont {Geballe}, \citenamefont {Matthias}, \citenamefont {Andres},
  \citenamefont {Schmidt},\ and\ \citenamefont
  {MacNair}}]{hannay1965superconductivity}%
  \BibitemOpen
  \bibfield  {author} {\bibinfo {author} {\bibfnamefont {N.}~\bibnamefont
  {Hannay}}, \bibinfo {author} {\bibfnamefont {T.}~\bibnamefont {Geballe}},
  \bibinfo {author} {\bibfnamefont {B.}~\bibnamefont {Matthias}}, \bibinfo
  {author} {\bibfnamefont {K.}~\bibnamefont {Andres}}, \bibinfo {author}
  {\bibfnamefont {P.}~\bibnamefont {Schmidt}}, \ and\ \bibinfo {author}
  {\bibfnamefont {D.}~\bibnamefont {MacNair}},\ }\href@noop {} {\bibfield
  {journal} {\bibinfo  {journal} {Physical Review Letters}\ }\textbf {\bibinfo
  {volume} {14}},\ \bibinfo {pages} {225} (\bibinfo {year} {1965})}\BibitemShut
  {NoStop}%
\bibitem [{\citenamefont {Belash}\ \emph {et~al.}(1989)\citenamefont {Belash},
  \citenamefont {Bronnikov}, \citenamefont {Zharikov},\ and\ \citenamefont
  {Pal'nichenko}}]{belash1989superconductivity}%
  \BibitemOpen
  \bibfield  {author} {\bibinfo {author} {\bibfnamefont {I.}~\bibnamefont
  {Belash}}, \bibinfo {author} {\bibfnamefont {A.}~\bibnamefont {Bronnikov}},
  \bibinfo {author} {\bibfnamefont {O.}~\bibnamefont {Zharikov}}, \ and\
  \bibinfo {author} {\bibfnamefont {A.}~\bibnamefont {Pal'nichenko}},\
  }\href@noop {} {\bibfield  {journal} {\bibinfo  {journal} {Solid State
  Communications}\ }\textbf {\bibinfo {volume} {69}},\ \bibinfo {pages} {921}
  (\bibinfo {year} {1989})}\BibitemShut {NoStop}%
\bibitem [{\citenamefont {Emery}\ \emph {et~al.}(2005)\citenamefont {Emery},
  \citenamefont {H{\'e}rold}, \citenamefont {d'Astuto}, \citenamefont {Garcia},
  \citenamefont {Bellin}, \citenamefont {Mar{\^e}ch{\'e}}, \citenamefont
  {Lagrange},\ and\ \citenamefont {Loupias}}]{emery2005superconductivity}%
  \BibitemOpen
  \bibfield  {author} {\bibinfo {author} {\bibfnamefont {N.}~\bibnamefont
  {Emery}}, \bibinfo {author} {\bibfnamefont {C.}~\bibnamefont {H{\'e}rold}},
  \bibinfo {author} {\bibfnamefont {M.}~\bibnamefont {d'Astuto}}, \bibinfo
  {author} {\bibfnamefont {V.}~\bibnamefont {Garcia}}, \bibinfo {author}
  {\bibfnamefont {C.}~\bibnamefont {Bellin}}, \bibinfo {author} {\bibfnamefont
  {J.}~\bibnamefont {Mar{\^e}ch{\'e}}}, \bibinfo {author} {\bibfnamefont
  {P.}~\bibnamefont {Lagrange}}, \ and\ \bibinfo {author} {\bibfnamefont
  {G.}~\bibnamefont {Loupias}},\ }\href@noop {} {\bibfield  {journal} {\bibinfo
   {journal} {Physical Review Letters}\ }\textbf {\bibinfo {volume} {95}},\
  \bibinfo {pages} {087003} (\bibinfo {year} {2005})}\BibitemShut {NoStop}%
\bibitem [{\citenamefont {Calandra}\ and\ \citenamefont
  {Mauri}(2005)}]{calandra2005theoretical}%
  \BibitemOpen
  \bibfield  {author} {\bibinfo {author} {\bibfnamefont {M.}~\bibnamefont
  {Calandra}}\ and\ \bibinfo {author} {\bibfnamefont {F.}~\bibnamefont
  {Mauri}},\ }\href@noop {} {\bibfield  {journal} {\bibinfo  {journal}
  {Physical Review Letters}\ }\textbf {\bibinfo {volume} {95}},\ \bibinfo
  {pages} {237002} (\bibinfo {year} {2005})}\BibitemShut {NoStop}%
\bibitem [{\citenamefont {Cs{\'a}nyi}\ \emph {et~al.}(2005)\citenamefont
  {Cs{\'a}nyi}, \citenamefont {Littlewood}, \citenamefont {Nevidomskyy},
  \citenamefont {Pickard},\ and\ \citenamefont {Simons}}]{csanyi2005role}%
  \BibitemOpen
  \bibfield  {author} {\bibinfo {author} {\bibfnamefont {G.}~\bibnamefont
  {Cs{\'a}nyi}}, \bibinfo {author} {\bibfnamefont {P.}~\bibnamefont
  {Littlewood}}, \bibinfo {author} {\bibfnamefont {A.~H.}\ \bibnamefont
  {Nevidomskyy}}, \bibinfo {author} {\bibfnamefont {C.~J.}\ \bibnamefont
  {Pickard}}, \ and\ \bibinfo {author} {\bibfnamefont {B.}~\bibnamefont
  {Simons}},\ }\href@noop {} {\bibfield  {journal} {\bibinfo  {journal} {Nature
  Physics}\ }\textbf {\bibinfo {volume} {1}},\ \bibinfo {pages} {42} (\bibinfo
  {year} {2005})}\BibitemShut {NoStop}%
\bibitem [{\citenamefont {Profeta}\ \emph {et~al.}(2012)\citenamefont
  {Profeta}, \citenamefont {Calandra},\ and\ \citenamefont
  {Mauri}}]{profeta2012phonon}%
  \BibitemOpen
  \bibfield  {author} {\bibinfo {author} {\bibfnamefont {G.}~\bibnamefont
  {Profeta}}, \bibinfo {author} {\bibfnamefont {M.}~\bibnamefont {Calandra}}, \
  and\ \bibinfo {author} {\bibfnamefont {F.}~\bibnamefont {Mauri}},\
  }\href@noop {} {\bibfield  {journal} {\bibinfo  {journal} {Nature Physics}\
  }\textbf {\bibinfo {volume} {8}},\ \bibinfo {pages} {131} (\bibinfo {year}
  {2012})}\BibitemShut {NoStop}%
\bibitem [{\citenamefont {Sanna}\ \emph {et~al.}(2007)\citenamefont {Sanna},
  \citenamefont {Profeta}, \citenamefont {Floris}, \citenamefont {Marini},
  \citenamefont {Gross},\ and\ \citenamefont
  {Massidda}}]{sanna2007anisotropic}%
  \BibitemOpen
  \bibfield  {author} {\bibinfo {author} {\bibfnamefont {A.}~\bibnamefont
  {Sanna}}, \bibinfo {author} {\bibfnamefont {G.}~\bibnamefont {Profeta}},
  \bibinfo {author} {\bibfnamefont {A.}~\bibnamefont {Floris}}, \bibinfo
  {author} {\bibfnamefont {A.}~\bibnamefont {Marini}}, \bibinfo {author}
  {\bibfnamefont {E.}~\bibnamefont {Gross}}, \ and\ \bibinfo {author}
  {\bibfnamefont {S.}~\bibnamefont {Massidda}},\ }\href@noop {} {\bibfield
  {journal} {\bibinfo  {journal} {Physical Review B}\ }\textbf {\bibinfo
  {volume} {75}},\ \bibinfo {pages} {020511} (\bibinfo {year}
  {2007})}\BibitemShut {NoStop}%
\bibitem [{\citenamefont {Gauzzi}\ \emph {et~al.}(2007)\citenamefont {Gauzzi},
  \citenamefont {Takashima}, \citenamefont {Takeshita}, \citenamefont
  {Terakura}, \citenamefont {Takagi}, \citenamefont {Emery}, \citenamefont
  {H{\'e}rold}, \citenamefont {Lagrange},\ and\ \citenamefont
  {Loupias}}]{gauzzi2007enhancement}%
  \BibitemOpen
  \bibfield  {author} {\bibinfo {author} {\bibfnamefont {A.}~\bibnamefont
  {Gauzzi}}, \bibinfo {author} {\bibfnamefont {S.}~\bibnamefont {Takashima}},
  \bibinfo {author} {\bibfnamefont {N.}~\bibnamefont {Takeshita}}, \bibinfo
  {author} {\bibfnamefont {C.}~\bibnamefont {Terakura}}, \bibinfo {author}
  {\bibfnamefont {H.}~\bibnamefont {Takagi}}, \bibinfo {author} {\bibfnamefont
  {N.}~\bibnamefont {Emery}}, \bibinfo {author} {\bibfnamefont
  {C.}~\bibnamefont {H{\'e}rold}}, \bibinfo {author} {\bibfnamefont
  {P.}~\bibnamefont {Lagrange}}, \ and\ \bibinfo {author} {\bibfnamefont
  {G.}~\bibnamefont {Loupias}},\ }\href@noop {} {\bibfield  {journal} {\bibinfo
   {journal} {Physical Review Letters}\ }\textbf {\bibinfo {volume} {98}},\
  \bibinfo {pages} {067002} (\bibinfo {year} {2007})}\BibitemShut {NoStop}%
\bibitem [{\citenamefont {Tiwari}\ \emph {et~al.}(2017)\citenamefont {Tiwari},
  \citenamefont {Shin}, \citenamefont {Hwang}, \citenamefont {Jung},
  \citenamefont {Park},\ and\ \citenamefont
  {Lee}}]{tiwari2017superconductivity}%
  \BibitemOpen
  \bibfield  {author} {\bibinfo {author} {\bibfnamefont {A.~P.}\ \bibnamefont
  {Tiwari}}, \bibinfo {author} {\bibfnamefont {S.}~\bibnamefont {Shin}},
  \bibinfo {author} {\bibfnamefont {E.}~\bibnamefont {Hwang}}, \bibinfo
  {author} {\bibfnamefont {S.-G.}\ \bibnamefont {Jung}}, \bibinfo {author}
  {\bibfnamefont {T.}~\bibnamefont {Park}}, \ and\ \bibinfo {author}
  {\bibfnamefont {H.}~\bibnamefont {Lee}},\ }\href@noop {} {\bibfield
  {journal} {\bibinfo  {journal} {Journal of Physics: Condensed Matter}\
  }\textbf {\bibinfo {volume} {29}},\ \bibinfo {pages} {445701} (\bibinfo
  {year} {2017})}\BibitemShut {NoStop}%
\bibitem [{\citenamefont {Nishiyama}\ \emph {et~al.}(2017)\citenamefont
  {Nishiyama}, \citenamefont {Fujita}, \citenamefont {Hoshi}, \citenamefont
  {Miao}, \citenamefont {Terao}, \citenamefont {Yang}, \citenamefont
  {Miyazaki}, \citenamefont {Goto}, \citenamefont {Kagayama}, \citenamefont
  {Shimizu} \emph {et~al.}}]{nishiyama2017preparation}%
  \BibitemOpen
  \bibfield  {author} {\bibinfo {author} {\bibfnamefont {S.}~\bibnamefont
  {Nishiyama}}, \bibinfo {author} {\bibfnamefont {H.}~\bibnamefont {Fujita}},
  \bibinfo {author} {\bibfnamefont {M.}~\bibnamefont {Hoshi}}, \bibinfo
  {author} {\bibfnamefont {X.}~\bibnamefont {Miao}}, \bibinfo {author}
  {\bibfnamefont {T.}~\bibnamefont {Terao}}, \bibinfo {author} {\bibfnamefont
  {X.}~\bibnamefont {Yang}}, \bibinfo {author} {\bibfnamefont {T.}~\bibnamefont
  {Miyazaki}}, \bibinfo {author} {\bibfnamefont {H.}~\bibnamefont {Goto}},
  \bibinfo {author} {\bibfnamefont {T.}~\bibnamefont {Kagayama}}, \bibinfo
  {author} {\bibfnamefont {K.}~\bibnamefont {Shimizu}},  \emph {et~al.},\
  }\href@noop {} {\bibfield  {journal} {\bibinfo  {journal} {Scientific
  Reports}\ }\textbf {\bibinfo {volume} {7}},\ \bibinfo {pages} {7436}
  (\bibinfo {year} {2017})}\BibitemShut {NoStop}%
\bibitem [{\citenamefont {Liao}\ \emph {et~al.}(2017)\citenamefont {Liao},
  \citenamefont {Zhao}, \citenamefont {Zhao}, \citenamefont {Xu},\ and\
  \citenamefont {Yang}}]{liao2017phonon}%
  \BibitemOpen
  \bibfield  {author} {\bibinfo {author} {\bibfnamefont {J.-H.}\ \bibnamefont
  {Liao}}, \bibinfo {author} {\bibfnamefont {Y.-C.}\ \bibnamefont {Zhao}},
  \bibinfo {author} {\bibfnamefont {Y.-J.}\ \bibnamefont {Zhao}}, \bibinfo
  {author} {\bibfnamefont {H.}~\bibnamefont {Xu}}, \ and\ \bibinfo {author}
  {\bibfnamefont {X.-B.}\ \bibnamefont {Yang}},\ }\href@noop {} {\bibfield
  {journal} {\bibinfo  {journal} {Physical Chemistry Chemical Physics}\
  }\textbf {\bibinfo {volume} {19}},\ \bibinfo {pages} {29237} (\bibinfo {year}
  {2017})}\BibitemShut {NoStop}%
\bibitem [{\citenamefont {Tawfik}\ \emph {et~al.}(2018)\citenamefont {Tawfik},
  \citenamefont {Stampfl},\ and\ \citenamefont {Ford}}]{Tawfik}%
  \BibitemOpen
  \bibfield  {author} {\bibinfo {author} {\bibfnamefont {S.~A.}\ \bibnamefont
  {Tawfik}}, \bibinfo {author} {\bibfnamefont {C.}~\bibnamefont {Stampfl}}, \
  and\ \bibinfo {author} {\bibfnamefont {M.~J.}\ \bibnamefont {Ford}},\
  }\href@noop {} {\bibfield  {journal} {\bibinfo  {journal} {Physical Chemistry
  Chemical Physics}\ }\textbf {\bibinfo {volume} {20}},\ \bibinfo {pages}
  {24027} (\bibinfo {year} {2018})}\BibitemShut {NoStop}%
\bibitem [{\citenamefont {Bharathi}\ \emph {et~al.}(2003)\citenamefont
  {Bharathi}, \citenamefont {Hariharan}, \citenamefont {Balaselvi},\ and\
  \citenamefont {Sundar}}]{bharathi2003superconductivity}%
  \BibitemOpen
  \bibfield  {author} {\bibinfo {author} {\bibfnamefont {A.}~\bibnamefont
  {Bharathi}}, \bibinfo {author} {\bibfnamefont {Y.}~\bibnamefont {Hariharan}},
  \bibinfo {author} {\bibfnamefont {J.}~\bibnamefont {Balaselvi}}, \ and\
  \bibinfo {author} {\bibfnamefont {C.}~\bibnamefont {Sundar}},\ }\href@noop {}
  {\bibfield  {journal} {\bibinfo  {journal} {Sadhana}\ }\textbf {\bibinfo
  {volume} {28}},\ \bibinfo {pages} {263} (\bibinfo {year} {2003})}\BibitemShut
  {NoStop}%
\bibitem [{\citenamefont {Jemima}\ \emph {et~al.}(2003)\citenamefont {Jemima},
  \citenamefont {Bharathi}, \citenamefont {Sankara}, \citenamefont {Reddy},
  \citenamefont {Hariharan} \emph {et~al.}}]{jemima2003effect}%
  \BibitemOpen
  \bibfield  {author} {\bibinfo {author} {\bibfnamefont {B.}~\bibnamefont
  {Jemima}}, \bibinfo {author} {\bibfnamefont {A.}~\bibnamefont {Bharathi}},
  \bibinfo {author} {\bibfnamefont {S.}~\bibnamefont {Sankara}}, \bibinfo
  {author} {\bibfnamefont {G.}~\bibnamefont {Reddy}}, \bibinfo {author}
  {\bibfnamefont {Y.}~\bibnamefont {Hariharan}},  \emph {et~al.},\ }in\
  \href@noop {} {\emph {\bibinfo {booktitle} {Proceedings of the DAE solid
  state physics symposium. V. 45}}}\ (\bibinfo {year} {2003})\BibitemShut
  {NoStop}%
\bibitem [{\citenamefont {Kazakov}\ \emph {et~al.}(2005)\citenamefont
  {Kazakov}, \citenamefont {Puzniak}, \citenamefont {Rogacki}, \citenamefont
  {Mironov}, \citenamefont {Zhigadlo}, \citenamefont {Jun}, \citenamefont
  {Soltmann}, \citenamefont {Batlogg},\ and\ \citenamefont
  {Karpinski}}]{kazakov2005carbon}%
  \BibitemOpen
  \bibfield  {author} {\bibinfo {author} {\bibfnamefont {S.}~\bibnamefont
  {Kazakov}}, \bibinfo {author} {\bibfnamefont {R.}~\bibnamefont {Puzniak}},
  \bibinfo {author} {\bibfnamefont {K.}~\bibnamefont {Rogacki}}, \bibinfo
  {author} {\bibfnamefont {A.}~\bibnamefont {Mironov}}, \bibinfo {author}
  {\bibfnamefont {N.}~\bibnamefont {Zhigadlo}}, \bibinfo {author}
  {\bibfnamefont {J.}~\bibnamefont {Jun}}, \bibinfo {author} {\bibfnamefont
  {C.}~\bibnamefont {Soltmann}}, \bibinfo {author} {\bibfnamefont
  {B.}~\bibnamefont {Batlogg}}, \ and\ \bibinfo {author} {\bibfnamefont
  {J.}~\bibnamefont {Karpinski}},\ }\href@noop {} {\bibfield  {journal}
  {\bibinfo  {journal} {Physical Review B}\ }\textbf {\bibinfo {volume} {71}},\
  \bibinfo {pages} {024533} (\bibinfo {year} {2005})}\BibitemShut {NoStop}%
\bibitem [{\citenamefont {Balaselvi}\ \emph {et~al.}(2004)\citenamefont
  {Balaselvi}, \citenamefont {Gayathri}, \citenamefont {Bharathi},
  \citenamefont {Sastry},\ and\ \citenamefont
  {Hariharan}}]{balaselvi2004stoichiometric}%
  \BibitemOpen
  \bibfield  {author} {\bibinfo {author} {\bibfnamefont {S.~J.}\ \bibnamefont
  {Balaselvi}}, \bibinfo {author} {\bibfnamefont {N.}~\bibnamefont {Gayathri}},
  \bibinfo {author} {\bibfnamefont {A.}~\bibnamefont {Bharathi}}, \bibinfo
  {author} {\bibfnamefont {V.}~\bibnamefont {Sastry}}, \ and\ \bibinfo {author}
  {\bibfnamefont {Y.}~\bibnamefont {Hariharan}},\ }\href@noop {} {\bibfield
  {journal} {\bibinfo  {journal} {Superconductor Science and Technology}\
  }\textbf {\bibinfo {volume} {17}},\ \bibinfo {pages} {1401} (\bibinfo {year}
  {2004})}\BibitemShut {NoStop}%
\bibitem [{\citenamefont {Ohmichi}\ \emph {et~al.}(2004)\citenamefont
  {Ohmichi}, \citenamefont {Masui}, \citenamefont {Lee}, \citenamefont
  {Tajima},\ and\ \citenamefont {Osada}}]{ohmichi2004enhancement}%
  \BibitemOpen
  \bibfield  {author} {\bibinfo {author} {\bibfnamefont {E.}~\bibnamefont
  {Ohmichi}}, \bibinfo {author} {\bibfnamefont {T.}~\bibnamefont {Masui}},
  \bibinfo {author} {\bibfnamefont {S.}~\bibnamefont {Lee}}, \bibinfo {author}
  {\bibfnamefont {S.}~\bibnamefont {Tajima}}, \ and\ \bibinfo {author}
  {\bibfnamefont {T.}~\bibnamefont {Osada}},\ }\href@noop {} {\bibfield
  {journal} {\bibinfo  {journal} {Journal of the Physical Society of Japan}\
  }\textbf {\bibinfo {volume} {73}},\ \bibinfo {pages} {2065} (\bibinfo {year}
  {2004})}\BibitemShut {NoStop}%
\bibitem [{\citenamefont {Bharathi}\ \emph
  {et~al.}(2002{\natexlab{a}})\citenamefont {Bharathi}, \citenamefont
  {Balaselvi}, \citenamefont {Kalavathi}, \citenamefont {Reddy}, \citenamefont
  {Sastry}, \citenamefont {Hariharan},\ and\ \citenamefont
  {Radhakrishnan}}]{bharathi2002carbon}%
  \BibitemOpen
  \bibfield  {author} {\bibinfo {author} {\bibfnamefont {A.}~\bibnamefont
  {Bharathi}}, \bibinfo {author} {\bibfnamefont {S.~J.}\ \bibnamefont
  {Balaselvi}}, \bibinfo {author} {\bibfnamefont {S.}~\bibnamefont
  {Kalavathi}}, \bibinfo {author} {\bibfnamefont {G.}~\bibnamefont {Reddy}},
  \bibinfo {author} {\bibfnamefont {V.~S.}\ \bibnamefont {Sastry}}, \bibinfo
  {author} {\bibfnamefont {Y.}~\bibnamefont {Hariharan}}, \ and\ \bibinfo
  {author} {\bibfnamefont {T.}~\bibnamefont {Radhakrishnan}},\ }\href@noop {}
  {\bibfield  {journal} {\bibinfo  {journal} {Physica C: Superconductivity}\
  }\textbf {\bibinfo {volume} {370}},\ \bibinfo {pages} {211} (\bibinfo {year}
  {2002}{\natexlab{a}})}\BibitemShut {NoStop}%
\bibitem [{\citenamefont {Braccini}\ \emph {et~al.}(2005)\citenamefont
  {Braccini}, \citenamefont {Gurevich}, \citenamefont {Giencke}, \citenamefont
  {Jewell}, \citenamefont {Eom}, \citenamefont {Larbalestier}, \citenamefont
  {Pogrebnyakov}, \citenamefont {Cui}, \citenamefont {Liu}, \citenamefont {Hu}
  \emph {et~al.}}]{braccini2005high}%
  \BibitemOpen
  \bibfield  {author} {\bibinfo {author} {\bibfnamefont {V.}~\bibnamefont
  {Braccini}}, \bibinfo {author} {\bibfnamefont {A.}~\bibnamefont {Gurevich}},
  \bibinfo {author} {\bibfnamefont {J.}~\bibnamefont {Giencke}}, \bibinfo
  {author} {\bibfnamefont {M.}~\bibnamefont {Jewell}}, \bibinfo {author}
  {\bibfnamefont {C.}~\bibnamefont {Eom}}, \bibinfo {author} {\bibfnamefont
  {D.}~\bibnamefont {Larbalestier}}, \bibinfo {author} {\bibfnamefont
  {A.}~\bibnamefont {Pogrebnyakov}}, \bibinfo {author} {\bibfnamefont
  {Y.}~\bibnamefont {Cui}}, \bibinfo {author} {\bibfnamefont {B.}~\bibnamefont
  {Liu}}, \bibinfo {author} {\bibfnamefont {Y.}~\bibnamefont {Hu}},  \emph
  {et~al.},\ }\href@noop {} {\bibfield  {journal} {\bibinfo  {journal}
  {Physical Review B}\ }\textbf {\bibinfo {volume} {71}},\ \bibinfo {pages}
  {012504} (\bibinfo {year} {2005})}\BibitemShut {NoStop}%
\bibitem [{\citenamefont {Pissas}\ \emph {et~al.}(2004)\citenamefont {Pissas},
  \citenamefont {Stamopoulos}, \citenamefont {Lee},\ and\ \citenamefont
  {Tajima}}]{pissas2004vortex}%
  \BibitemOpen
  \bibfield  {author} {\bibinfo {author} {\bibfnamefont {M.}~\bibnamefont
  {Pissas}}, \bibinfo {author} {\bibfnamefont {D.}~\bibnamefont {Stamopoulos}},
  \bibinfo {author} {\bibfnamefont {S.}~\bibnamefont {Lee}}, \ and\ \bibinfo
  {author} {\bibfnamefont {S.}~\bibnamefont {Tajima}},\ }\href@noop {}
  {\bibfield  {journal} {\bibinfo  {journal} {Physical Review B}\ }\textbf
  {\bibinfo {volume} {70}},\ \bibinfo {pages} {134503} (\bibinfo {year}
  {2004})}\BibitemShut {NoStop}%
\bibitem [{\citenamefont {Rosner}\ \emph {et~al.}(2002)\citenamefont {Rosner},
  \citenamefont {Kitaigorodsky},\ and\ \citenamefont
  {Pickett}}]{rosner2002prediction}%
  \BibitemOpen
  \bibfield  {author} {\bibinfo {author} {\bibfnamefont {H.}~\bibnamefont
  {Rosner}}, \bibinfo {author} {\bibfnamefont {A.}~\bibnamefont
  {Kitaigorodsky}}, \ and\ \bibinfo {author} {\bibfnamefont {W.}~\bibnamefont
  {Pickett}},\ }\href@noop {} {\bibfield  {journal} {\bibinfo  {journal}
  {Physical Review Letters}\ }\textbf {\bibinfo {volume} {88}},\ \bibinfo
  {pages} {127001} (\bibinfo {year} {2002})}\BibitemShut {NoStop}%
\bibitem [{\citenamefont {Fogg}\ \emph {et~al.}(2003)\citenamefont {Fogg},
  \citenamefont {Chalker}, \citenamefont {Claridge}, \citenamefont {Darling},\
  and\ \citenamefont {Rosseinsky}}]{fogg2003libc}%
  \BibitemOpen
  \bibfield  {author} {\bibinfo {author} {\bibfnamefont {A.}~\bibnamefont
  {Fogg}}, \bibinfo {author} {\bibfnamefont {P.}~\bibnamefont {Chalker}},
  \bibinfo {author} {\bibfnamefont {J.}~\bibnamefont {Claridge}}, \bibinfo
  {author} {\bibfnamefont {G.}~\bibnamefont {Darling}}, \ and\ \bibinfo
  {author} {\bibfnamefont {M.}~\bibnamefont {Rosseinsky}},\ }\href@noop {}
  {\bibfield  {journal} {\bibinfo  {journal} {Physical Review B}\ }\textbf
  {\bibinfo {volume} {67}},\ \bibinfo {pages} {245106} (\bibinfo {year}
  {2003})}\BibitemShut {NoStop}%
\bibitem [{\citenamefont {Fogg}\ \emph {et~al.}(2006)\citenamefont {Fogg},
  \citenamefont {Meldrum}, \citenamefont {Darling}, \citenamefont {Claridge},\
  and\ \citenamefont {Rosseinsky}}]{fogg2006chemical}%
  \BibitemOpen
  \bibfield  {author} {\bibinfo {author} {\bibfnamefont {A.~M.}\ \bibnamefont
  {Fogg}}, \bibinfo {author} {\bibfnamefont {J.}~\bibnamefont {Meldrum}},
  \bibinfo {author} {\bibfnamefont {G.~R.}\ \bibnamefont {Darling}}, \bibinfo
  {author} {\bibfnamefont {J.~B.}\ \bibnamefont {Claridge}}, \ and\ \bibinfo
  {author} {\bibfnamefont {M.~J.}\ \bibnamefont {Rosseinsky}},\ }\href@noop {}
  {\bibfield  {journal} {\bibinfo  {journal} {Journal of the American Chemical
  Society}\ }\textbf {\bibinfo {volume} {128}},\ \bibinfo {pages} {10043}
  (\bibinfo {year} {2006})}\BibitemShut {NoStop}%
\bibitem [{\citenamefont {Souptel}\ \emph {et~al.}(2003)\citenamefont
  {Souptel}, \citenamefont {Hossain}, \citenamefont {Behr}, \citenamefont
  {L{\"o}ser},\ and\ \citenamefont {Geibel}}]{souptel2003synthesis}%
  \BibitemOpen
  \bibfield  {author} {\bibinfo {author} {\bibfnamefont {D.}~\bibnamefont
  {Souptel}}, \bibinfo {author} {\bibfnamefont {Z.}~\bibnamefont {Hossain}},
  \bibinfo {author} {\bibfnamefont {G.}~\bibnamefont {Behr}}, \bibinfo {author}
  {\bibfnamefont {W.}~\bibnamefont {L{\"o}ser}}, \ and\ \bibinfo {author}
  {\bibfnamefont {C.}~\bibnamefont {Geibel}},\ }\href@noop {} {\bibfield
  {journal} {\bibinfo  {journal} {Solid State Communications}\ }\textbf
  {\bibinfo {volume} {125}},\ \bibinfo {pages} {17} (\bibinfo {year}
  {2003})}\BibitemShut {NoStop}%
\bibitem [{\citenamefont {Bharathi}\ \emph
  {et~al.}(2002{\natexlab{b}})\citenamefont {Bharathi}, \citenamefont
  {Balaselvi}, \citenamefont {Premila}, \citenamefont {Sairam}, \citenamefont
  {Reddy}, \citenamefont {Sundar},\ and\ \citenamefont
  {Hariharan}}]{bharathi2002synthesis}%
  \BibitemOpen
  \bibfield  {author} {\bibinfo {author} {\bibfnamefont {A.}~\bibnamefont
  {Bharathi}}, \bibinfo {author} {\bibfnamefont {S.~J.}\ \bibnamefont
  {Balaselvi}}, \bibinfo {author} {\bibfnamefont {M.}~\bibnamefont {Premila}},
  \bibinfo {author} {\bibfnamefont {T.}~\bibnamefont {Sairam}}, \bibinfo
  {author} {\bibfnamefont {G.}~\bibnamefont {Reddy}}, \bibinfo {author}
  {\bibfnamefont {C.}~\bibnamefont {Sundar}}, \ and\ \bibinfo {author}
  {\bibfnamefont {Y.}~\bibnamefont {Hariharan}},\ }\href@noop {} {\bibfield
  {journal} {\bibinfo  {journal} {Solid State Communications}\ }\textbf
  {\bibinfo {volume} {124}},\ \bibinfo {pages} {423} (\bibinfo {year}
  {2002}{\natexlab{b}})}\BibitemShut {NoStop}%
\bibitem [{\citenamefont {Gao}\ \emph {et~al.}(2015)\citenamefont {Gao},
  \citenamefont {Lu},\ and\ \citenamefont {Xiang}}]{gao2015prediction}%
  \BibitemOpen
  \bibfield  {author} {\bibinfo {author} {\bibfnamefont {M.}~\bibnamefont
  {Gao}}, \bibinfo {author} {\bibfnamefont {Z.-Y.}\ \bibnamefont {Lu}}, \ and\
  \bibinfo {author} {\bibfnamefont {T.}~\bibnamefont {Xiang}},\ }\href@noop {}
  {\bibfield  {journal} {\bibinfo  {journal} {Physical Review B}\ }\textbf
  {\bibinfo {volume} {91}},\ \bibinfo {pages} {045132} (\bibinfo {year}
  {2015})}\BibitemShut {NoStop}%
\bibitem [{\citenamefont {Li}\ \emph {et~al.}(2018)\citenamefont {Li},
  \citenamefont {Yan}, \citenamefont {Gao},\ and\ \citenamefont
  {Wang}}]{Qielectron}%
  \BibitemOpen
  \bibfield  {author} {\bibinfo {author} {\bibfnamefont {Q.-Z.}\ \bibnamefont
  {Li}}, \bibinfo {author} {\bibfnamefont {X.-W.}\ \bibnamefont {Yan}},
  \bibinfo {author} {\bibfnamefont {M.}~\bibnamefont {Gao}}, \ and\ \bibinfo
  {author} {\bibfnamefont {J.}~\bibnamefont {Wang}},\ }\href
  {http://stacks.iop.org/0295-5075/122/i=4/a=47001} {\bibfield  {journal}
  {\bibinfo  {journal} {Europhysics Letters}\ }\textbf {\bibinfo {volume}
  {122}},\ \bibinfo {pages} {47001} (\bibinfo {year} {2018})}\BibitemShut
  {NoStop}%
\bibitem [{\citenamefont {Bazhirov}\ \emph {et~al.}(2014)\citenamefont
  {Bazhirov}, \citenamefont {Sakai}, \citenamefont {Saito},\ and\ \citenamefont
  {Cohen}}]{bazhirov2014electron}%
  \BibitemOpen
  \bibfield  {author} {\bibinfo {author} {\bibfnamefont {T.}~\bibnamefont
  {Bazhirov}}, \bibinfo {author} {\bibfnamefont {Y.}~\bibnamefont {Sakai}},
  \bibinfo {author} {\bibfnamefont {S.}~\bibnamefont {Saito}}, \ and\ \bibinfo
  {author} {\bibfnamefont {M.~L.}\ \bibnamefont {Cohen}},\ }\href@noop {}
  {\bibfield  {journal} {\bibinfo  {journal} {Physical Review B}\ }\textbf
  {\bibinfo {volume} {89}},\ \bibinfo {pages} {045136} (\bibinfo {year}
  {2014})}\BibitemShut {NoStop}%
\bibitem [{\citenamefont {Miao}\ \emph {et~al.}(2016)\citenamefont {Miao},
  \citenamefont {Huang},\ and\ \citenamefont {Yang}}]{miao2016first}%
  \BibitemOpen
  \bibfield  {author} {\bibinfo {author} {\bibfnamefont {R.}~\bibnamefont
  {Miao}}, \bibinfo {author} {\bibfnamefont {G.}~\bibnamefont {Huang}}, \ and\
  \bibinfo {author} {\bibfnamefont {J.}~\bibnamefont {Yang}},\ }\href@noop {}
  {\bibfield  {journal} {\bibinfo  {journal} {Solid State Communications}\
  }\textbf {\bibinfo {volume} {233}},\ \bibinfo {pages} {30} (\bibinfo {year}
  {2016})}\BibitemShut {NoStop}%
\bibitem [{\citenamefont {Ravindran}\ \emph {et~al.}(2001)\citenamefont
  {Ravindran}, \citenamefont {Vajeeston}, \citenamefont {Vidya}, \citenamefont
  {Kjekshus},\ and\ \citenamefont {Fjellv{\aa}g}}]{ravindran2001detailed}%
  \BibitemOpen
  \bibfield  {author} {\bibinfo {author} {\bibfnamefont {P.}~\bibnamefont
  {Ravindran}}, \bibinfo {author} {\bibfnamefont {P.}~\bibnamefont
  {Vajeeston}}, \bibinfo {author} {\bibfnamefont {R.}~\bibnamefont {Vidya}},
  \bibinfo {author} {\bibfnamefont {A.}~\bibnamefont {Kjekshus}}, \ and\
  \bibinfo {author} {\bibfnamefont {H.}~\bibnamefont {Fjellv{\aa}g}},\
  }\href@noop {} {\bibfield  {journal} {\bibinfo  {journal} {Physical Review
  B}\ }\textbf {\bibinfo {volume} {64}},\ \bibinfo {pages} {224509} (\bibinfo
  {year} {2001})}\BibitemShut {NoStop}%
\bibitem [{\citenamefont {Perdew}\ \emph {et~al.}(1996)\citenamefont {Perdew},
  \citenamefont {Burke},\ and\ \citenamefont
  {Ernzerhof}}]{perdew1996generalized}%
  \BibitemOpen
  \bibfield  {author} {\bibinfo {author} {\bibfnamefont {J.~P.}\ \bibnamefont
  {Perdew}}, \bibinfo {author} {\bibfnamefont {K.}~\bibnamefont {Burke}}, \
  and\ \bibinfo {author} {\bibfnamefont {M.}~\bibnamefont {Ernzerhof}},\
  }\href@noop {} {\bibfield  {journal} {\bibinfo  {journal} {Physical Review
  Letters}\ }\textbf {\bibinfo {volume} {77}},\ \bibinfo {pages} {3865}
  (\bibinfo {year} {1996})}\BibitemShut {NoStop}%
\bibitem [{\citenamefont {Perdew}\ \emph {et~al.}(2008)\citenamefont {Perdew},
  \citenamefont {Ruzsinszky}, \citenamefont {Csonka}, \citenamefont {Vydrov},
  \citenamefont {Scuseria}, \citenamefont {Constantin}, \citenamefont {Zhou},\
  and\ \citenamefont {Burke}}]{perdew2008restoring}%
  \BibitemOpen
  \bibfield  {author} {\bibinfo {author} {\bibfnamefont {J.~P.}\ \bibnamefont
  {Perdew}}, \bibinfo {author} {\bibfnamefont {A.}~\bibnamefont {Ruzsinszky}},
  \bibinfo {author} {\bibfnamefont {G.~I.}\ \bibnamefont {Csonka}}, \bibinfo
  {author} {\bibfnamefont {O.~A.}\ \bibnamefont {Vydrov}}, \bibinfo {author}
  {\bibfnamefont {G.~E.}\ \bibnamefont {Scuseria}}, \bibinfo {author}
  {\bibfnamefont {L.~A.}\ \bibnamefont {Constantin}}, \bibinfo {author}
  {\bibfnamefont {X.}~\bibnamefont {Zhou}}, \ and\ \bibinfo {author}
  {\bibfnamefont {K.}~\bibnamefont {Burke}},\ }\href@noop {} {\bibfield
  {journal} {\bibinfo  {journal} {Physical Review Letters}\ }\textbf {\bibinfo
  {volume} {100}},\ \bibinfo {pages} {136406} (\bibinfo {year}
  {2008})}\BibitemShut {NoStop}%
\bibitem [{\citenamefont {Giannozzi}\ \emph {et~al.}(2009)\citenamefont
  {Giannozzi}, \citenamefont {Baroni}, \citenamefont {Bonini}, \citenamefont
  {Calandra}, \citenamefont {Car}, \citenamefont {Cavazzoni}, \citenamefont
  {Ceresoli}, \citenamefont {Chiarotti}, \citenamefont {Cococcioni},
  \citenamefont {Dabo} \emph {et~al.}}]{giannozzi2009quantum}%
  \BibitemOpen
  \bibfield  {author} {\bibinfo {author} {\bibfnamefont {P.}~\bibnamefont
  {Giannozzi}}, \bibinfo {author} {\bibfnamefont {S.}~\bibnamefont {Baroni}},
  \bibinfo {author} {\bibfnamefont {N.}~\bibnamefont {Bonini}}, \bibinfo
  {author} {\bibfnamefont {M.}~\bibnamefont {Calandra}}, \bibinfo {author}
  {\bibfnamefont {R.}~\bibnamefont {Car}}, \bibinfo {author} {\bibfnamefont
  {C.}~\bibnamefont {Cavazzoni}}, \bibinfo {author} {\bibfnamefont
  {D.}~\bibnamefont {Ceresoli}}, \bibinfo {author} {\bibfnamefont {G.~L.}\
  \bibnamefont {Chiarotti}}, \bibinfo {author} {\bibfnamefont {M.}~\bibnamefont
  {Cococcioni}}, \bibinfo {author} {\bibfnamefont {I.}~\bibnamefont {Dabo}},
  \emph {et~al.},\ }\href@noop {} {\bibfield  {journal} {\bibinfo  {journal}
  {Journal of Physics: Condensed Matter}\ }\textbf {\bibinfo {volume} {21}},\
  \bibinfo {pages} {395502} (\bibinfo {year} {2009})}\BibitemShut {NoStop}%
\bibitem [{\citenamefont {Vanderbilt}(1990)}]{vanderbilt1990soft}%
  \BibitemOpen
  \bibfield  {author} {\bibinfo {author} {\bibfnamefont {D.}~\bibnamefont
  {Vanderbilt}},\ }\href@noop {} {\bibfield  {journal} {\bibinfo  {journal}
  {Physical Review B}\ }\textbf {\bibinfo {volume} {41}},\ \bibinfo {pages}
  {7892} (\bibinfo {year} {1990})}\BibitemShut {NoStop}%
\bibitem [{tex()}]{textMgB2}%
  \BibitemOpen
  \href@noop {} {\ \textbf {\bibinfo {volume} {\normalfont We use same the
  cutoff energy and a 666 uniform grid of \textbf{q}-point for MgB$_2$. For
  calculation of the electron-phonon coupling constant and Fermi surface
  evaluation, we use a $24\times 24\times 24$ \textbf{k}-point
  mesh.}}}\BibitemShut {Stop}%
\bibitem [{\citenamefont {Kawamura}\ \emph {et~al.}(2014)\citenamefont
  {Kawamura}, \citenamefont {Gohda},\ and\ \citenamefont
  {Tsuneyuki}}]{kawamura2014improved}%
  \BibitemOpen
  \bibfield  {author} {\bibinfo {author} {\bibfnamefont {M.}~\bibnamefont
  {Kawamura}}, \bibinfo {author} {\bibfnamefont {Y.}~\bibnamefont {Gohda}}, \
  and\ \bibinfo {author} {\bibfnamefont {S.}~\bibnamefont {Tsuneyuki}},\
  }\href@noop {} {\bibfield  {journal} {\bibinfo  {journal} {Physical Review
  B}\ }\textbf {\bibinfo {volume} {89}},\ \bibinfo {pages} {094515} (\bibinfo
  {year} {2014})}\BibitemShut {NoStop}%
\bibitem [{\citenamefont {Baroni}\ \emph {et~al.}(2001)\citenamefont {Baroni},
  \citenamefont {De~Gironcoli}, \citenamefont {Dal~Corso},\ and\ \citenamefont
  {Giannozzi}}]{baroni2001phonons}%
  \BibitemOpen
  \bibfield  {author} {\bibinfo {author} {\bibfnamefont {S.}~\bibnamefont
  {Baroni}}, \bibinfo {author} {\bibfnamefont {S.}~\bibnamefont
  {De~Gironcoli}}, \bibinfo {author} {\bibfnamefont {A.}~\bibnamefont
  {Dal~Corso}}, \ and\ \bibinfo {author} {\bibfnamefont {P.}~\bibnamefont
  {Giannozzi}},\ }\href@noop {} {\bibfield  {journal} {\bibinfo  {journal}
  {Reviews of Modern Physics}\ }\textbf {\bibinfo {volume} {73}},\ \bibinfo
  {pages} {515} (\bibinfo {year} {2001})}\BibitemShut {NoStop}%
\bibitem [{\citenamefont {Eliashberg}(1960)}]{eliashberg1960interactions}%
  \BibitemOpen
  \bibfield  {author} {\bibinfo {author} {\bibfnamefont {G.}~\bibnamefont
  {Eliashberg}},\ }\href@noop {} {\bibfield  {journal} {\bibinfo  {journal}
  {Sov. Phys. JETP}\ }\textbf {\bibinfo {volume} {11}},\ \bibinfo {pages} {696}
  (\bibinfo {year} {1960})}\BibitemShut {NoStop}%
\bibitem [{\citenamefont {Allen}(1972)}]{allen1972neutron}%
  \BibitemOpen
  \bibfield  {author} {\bibinfo {author} {\bibfnamefont {P.~B.}\ \bibnamefont
  {Allen}},\ }\href@noop {} {\bibfield  {journal} {\bibinfo  {journal}
  {Physical Review B}\ }\textbf {\bibinfo {volume} {6}},\ \bibinfo {pages}
  {2577} (\bibinfo {year} {1972})}\BibitemShut {NoStop}%
\bibitem [{\citenamefont {Allen}\ and\ \citenamefont
  {Dynes}(1975)}]{allen1975transition}%
  \BibitemOpen
  \bibfield  {author} {\bibinfo {author} {\bibfnamefont {P.~B.}\ \bibnamefont
  {Allen}}\ and\ \bibinfo {author} {\bibfnamefont {R.}~\bibnamefont {Dynes}},\
  }\href@noop {} {\bibfield  {journal} {\bibinfo  {journal} {Physical Review
  B}\ }\textbf {\bibinfo {volume} {12}},\ \bibinfo {pages} {905} (\bibinfo
  {year} {1975})}\BibitemShut {NoStop}%
\bibitem [{\citenamefont {Richardson}\ and\ \citenamefont
  {Ashcroft}(1997)}]{richardson1997high}%
  \BibitemOpen
  \bibfield  {author} {\bibinfo {author} {\bibfnamefont {C.}~\bibnamefont
  {Richardson}}\ and\ \bibinfo {author} {\bibfnamefont {N.}~\bibnamefont
  {Ashcroft}},\ }\href@noop {} {\bibfield  {journal} {\bibinfo  {journal}
  {Physical Review Letters}\ }\textbf {\bibinfo {volume} {78}},\ \bibinfo
  {pages} {118} (\bibinfo {year} {1997})}\BibitemShut {NoStop}%
\bibitem [{\citenamefont {Lee}\ \emph {et~al.}(1995)\citenamefont {Lee},
  \citenamefont {Chang},\ and\ \citenamefont {Cohen}}]{lee1995first}%
  \BibitemOpen
  \bibfield  {author} {\bibinfo {author} {\bibfnamefont {K.-H.}\ \bibnamefont
  {Lee}}, \bibinfo {author} {\bibfnamefont {K.-J.}\ \bibnamefont {Chang}}, \
  and\ \bibinfo {author} {\bibfnamefont {M.~L.}\ \bibnamefont {Cohen}},\
  }\href@noop {} {\bibfield  {journal} {\bibinfo  {journal} {Physical Review
  B}\ }\textbf {\bibinfo {volume} {52}},\ \bibinfo {pages} {1425} (\bibinfo
  {year} {1995})}\BibitemShut {NoStop}%
\bibitem [{ena()}]{enamultext}%
  \BibitemOpen
  \href@noop {} {\ \textbf {\bibinfo {volume} {\normalfont The formation energy
  is calculated by taking the energy difference between the total energy of the
  compound and sum of the energy of individual constituent elements in their
  stable structure. We used hexagonal-Mg, face-centered cubic Ca and Sr,
  body-centered cubic Ba, graphite (C), and $\alpha$-$B_{12}$ type structure of
  boron \cite{decker1959crystal}. Our calculated formation energies of MgBC,
  CaBC, SrBC and BaBC are $-$0.14, $-$0.13, 0.73 and 1.5 eV.fu$^{-1}$,
  respectively. Although the formation energy of MgBC is small, the negative
  value of the formation energy of the first two compounds indicates a high
  possibility of synthesis of first two compounds in the laboratory. However,
  the positive formation energy of SrBC and BaBC indicates difficulty in the
  synthesis. The non-equilibrium methods of growth (epitaxial technique) may be
  used to synthesis these compounds (as a metastable structure), although it
  might be difficult}}}\BibitemShut {NoStop}%
\bibitem [{\citenamefont {Mazin}\ and\ \citenamefont
  {Antropov}(2003)}]{mazin2003electronic}%
  \BibitemOpen
  \bibfield  {author} {\bibinfo {author} {\bibfnamefont {I.}~\bibnamefont
  {Mazin}}\ and\ \bibinfo {author} {\bibfnamefont {V.}~\bibnamefont
  {Antropov}},\ }\href@noop {} {\bibfield  {journal} {\bibinfo  {journal}
  {Physica C: Superconductivity}\ }\textbf {\bibinfo {volume} {385}},\ \bibinfo
  {pages} {49} (\bibinfo {year} {2003})}\BibitemShut {NoStop}%
\bibitem [{\citenamefont {Grimvall}\ \emph {et~al.}(2012)\citenamefont
  {Grimvall}, \citenamefont {Magyari-K{\"o}pe}, \citenamefont
  {Ozoli{\c{n}}{\v{s}}},\ and\ \citenamefont {Persson}}]{grimvall2012lattice}%
  \BibitemOpen
  \bibfield  {author} {\bibinfo {author} {\bibfnamefont {G.}~\bibnamefont
  {Grimvall}}, \bibinfo {author} {\bibfnamefont {B.}~\bibnamefont
  {Magyari-K{\"o}pe}}, \bibinfo {author} {\bibfnamefont {V.}~\bibnamefont
  {Ozoli{\c{n}}{\v{s}}}}, \ and\ \bibinfo {author} {\bibfnamefont {K.~A.}\
  \bibnamefont {Persson}},\ }\href@noop {} {\bibfield  {journal} {\bibinfo
  {journal} {Reviews of Modern Physics}\ }\textbf {\bibinfo {volume} {84}},\
  \bibinfo {pages} {945} (\bibinfo {year} {2012})}\BibitemShut {NoStop}%
\bibitem [{\citenamefont {Kato}\ \emph {et~al.}(2004)\citenamefont {Kato},
  \citenamefont {Nagao}, \citenamefont {Nishikawa}, \citenamefont {Nishidate},\
  and\ \citenamefont {Endo}}]{kato2004possibility}%
  \BibitemOpen
  \bibfield  {author} {\bibinfo {author} {\bibfnamefont {N.}~\bibnamefont
  {Kato}}, \bibinfo {author} {\bibfnamefont {H.}~\bibnamefont {Nagao}},
  \bibinfo {author} {\bibfnamefont {K.}~\bibnamefont {Nishikawa}}, \bibinfo
  {author} {\bibfnamefont {K.}~\bibnamefont {Nishidate}}, \ and\ \bibinfo
  {author} {\bibfnamefont {K.}~\bibnamefont {Endo}},\ }\href@noop {} {\bibfield
   {journal} {\bibinfo  {journal} {International Journal of Quantum Chemistry}\
  }\textbf {\bibinfo {volume} {96}},\ \bibinfo {pages} {457} (\bibinfo {year}
  {2004})}\BibitemShut {NoStop}%
\bibitem [{lat()}]{latticesta}%
  \BibitemOpen
  \href@noop {} {\ \textbf {\bibinfo {volume} {\normalfont The large value of
  electron-phonon coupling of MgBC may give rise to the question of the lattice
  instability. All the predicted compounds have 18 independent vibrational
  modes. We have calculated the frequencies of each mode by using a uniform
  $q_1 \times q_1 \times q_2$ grid, where $q_1$ has been varied from 4 to 8 and
  $q_2$ from 2 to 4. We do not find any imaginary frequencies at these phonon
  wave vectors and along the considered high-symmetry \textbf{k}-points.
  Moreover, we have calculated the phonon dispersion using a Mazari Vanderbilt
  smearing of width 0.08 Ry and obtained the same phonon dispersion curves.
  Furthermore, we have calculated the phonon dispersion using the finite
  displacement method in the Phonopy program \cite{togo2015first} and obtained
  identical results. The obtained phonon dispersion curves suggest that $X$BC
  are dynamically stable.}}}\BibitemShut {Stop}%
\bibitem [{\citenamefont {Kong}\ \emph {et~al.}(2001)\citenamefont {Kong},
  \citenamefont {Dolgov}, \citenamefont {Jepsen},\ and\ \citenamefont
  {Andersen}}]{kong2001electron}%
  \BibitemOpen
  \bibfield  {author} {\bibinfo {author} {\bibfnamefont {Y.}~\bibnamefont
  {Kong}}, \bibinfo {author} {\bibfnamefont {O.}~\bibnamefont {Dolgov}},
  \bibinfo {author} {\bibfnamefont {O.}~\bibnamefont {Jepsen}}, \ and\ \bibinfo
  {author} {\bibfnamefont {O.}~\bibnamefont {Andersen}},\ }\href@noop {}
  {\bibfield  {journal} {\bibinfo  {journal} {Physical Review B}\ }\textbf
  {\bibinfo {volume} {64}},\ \bibinfo {pages} {020501} (\bibinfo {year}
  {2001})}\BibitemShut {NoStop}%
\bibitem [{\citenamefont {Yildirim}\ \emph {et~al.}(2001)\citenamefont
  {Yildirim}, \citenamefont {G{\"u}lseren}, \citenamefont {Lynn}, \citenamefont
  {Brown}, \citenamefont {Udovic}, \citenamefont {Huang}, \citenamefont
  {Rogado}, \citenamefont {Regan}, \citenamefont {Hayward}, \citenamefont
  {Slusky} \emph {et~al.}}]{yildirim2001giant}%
  \BibitemOpen
  \bibfield  {author} {\bibinfo {author} {\bibfnamefont {T.}~\bibnamefont
  {Yildirim}}, \bibinfo {author} {\bibfnamefont {O.}~\bibnamefont
  {G{\"u}lseren}}, \bibinfo {author} {\bibfnamefont {J.}~\bibnamefont {Lynn}},
  \bibinfo {author} {\bibfnamefont {C.}~\bibnamefont {Brown}}, \bibinfo
  {author} {\bibfnamefont {T.}~\bibnamefont {Udovic}}, \bibinfo {author}
  {\bibfnamefont {Q.}~\bibnamefont {Huang}}, \bibinfo {author} {\bibfnamefont
  {N.}~\bibnamefont {Rogado}}, \bibinfo {author} {\bibfnamefont
  {K.}~\bibnamefont {Regan}}, \bibinfo {author} {\bibfnamefont
  {M.}~\bibnamefont {Hayward}}, \bibinfo {author} {\bibfnamefont
  {J.}~\bibnamefont {Slusky}},  \emph {et~al.},\ }\href@noop {} {\bibfield
  {journal} {\bibinfo  {journal} {Physical Review Letters}\ }\textbf {\bibinfo
  {volume} {87}},\ \bibinfo {pages} {037001} (\bibinfo {year}
  {2001})}\BibitemShut {NoStop}%
\bibitem [{\citenamefont {Eiguren}\ and\ \citenamefont
  {Ambrosch-Draxl}(2008)}]{eiguren2008wannier}%
  \BibitemOpen
  \bibfield  {author} {\bibinfo {author} {\bibfnamefont {A.}~\bibnamefont
  {Eiguren}}\ and\ \bibinfo {author} {\bibfnamefont {C.}~\bibnamefont
  {Ambrosch-Draxl}},\ }\href@noop {} {\bibfield  {journal} {\bibinfo  {journal}
  {Physical Review B}\ }\textbf {\bibinfo {volume} {78}},\ \bibinfo {pages}
  {045124} (\bibinfo {year} {2008})}\BibitemShut {NoStop}%
\bibitem [{\citenamefont {Calandra}\ \emph {et~al.}(2010)\citenamefont
  {Calandra}, \citenamefont {Profeta},\ and\ \citenamefont
  {Mauri}}]{calandra2010adiabatic}%
  \BibitemOpen
  \bibfield  {author} {\bibinfo {author} {\bibfnamefont {M.}~\bibnamefont
  {Calandra}}, \bibinfo {author} {\bibfnamefont {G.}~\bibnamefont {Profeta}}, \
  and\ \bibinfo {author} {\bibfnamefont {F.}~\bibnamefont {Mauri}},\
  }\href@noop {} {\bibfield  {journal} {\bibinfo  {journal} {Physical Review
  B}\ }\textbf {\bibinfo {volume} {82}},\ \bibinfo {pages} {165111} (\bibinfo
  {year} {2010})}\BibitemShut {NoStop}%
\bibitem [{\citenamefont {Margine}\ and\ \citenamefont
  {Giustino}(2013)}]{margine2013anisotropic}%
  \BibitemOpen
  \bibfield  {author} {\bibinfo {author} {\bibfnamefont {E.~R.}\ \bibnamefont
  {Margine}}\ and\ \bibinfo {author} {\bibfnamefont {F.}~\bibnamefont
  {Giustino}},\ }\href@noop {} {\bibfield  {journal} {\bibinfo  {journal}
  {Physical Review B}\ }\textbf {\bibinfo {volume} {87}},\ \bibinfo {pages}
  {024505} (\bibinfo {year} {2013})}\BibitemShut {NoStop}%
\bibitem [{\citenamefont {An}\ and\ \citenamefont
  {Pickett}(2001)}]{an2001superconductivity}%
  \BibitemOpen
  \bibfield  {author} {\bibinfo {author} {\bibfnamefont {J.}~\bibnamefont
  {An}}\ and\ \bibinfo {author} {\bibfnamefont {W.}~\bibnamefont {Pickett}},\
  }\href@noop {} {\bibfield  {journal} {\bibinfo  {journal} {Physical Review
  Letters}\ }\textbf {\bibinfo {volume} {86}},\ \bibinfo {pages} {4366}
  (\bibinfo {year} {2001})}\BibitemShut {NoStop}%
\bibitem [{\citenamefont {Choi}\ \emph
  {et~al.}(2002{\natexlab{a}})\citenamefont {Choi}, \citenamefont {Roundy},
  \citenamefont {Sun}, \citenamefont {Cohen},\ and\ \citenamefont
  {Louie}}]{choi2002first}%
  \BibitemOpen
  \bibfield  {author} {\bibinfo {author} {\bibfnamefont {H.~J.}\ \bibnamefont
  {Choi}}, \bibinfo {author} {\bibfnamefont {D.}~\bibnamefont {Roundy}},
  \bibinfo {author} {\bibfnamefont {H.}~\bibnamefont {Sun}}, \bibinfo {author}
  {\bibfnamefont {M.~L.}\ \bibnamefont {Cohen}}, \ and\ \bibinfo {author}
  {\bibfnamefont {S.~G.}\ \bibnamefont {Louie}},\ }\href@noop {} {\bibfield
  {journal} {\bibinfo  {journal} {Physical Review B}\ }\textbf {\bibinfo
  {volume} {66}},\ \bibinfo {pages} {020513} (\bibinfo {year}
  {2002}{\natexlab{a}})}\BibitemShut {NoStop}%
\bibitem [{\citenamefont {Choi}\ \emph
  {et~al.}(2002{\natexlab{b}})\citenamefont {Choi}, \citenamefont {Roundy},
  \citenamefont {Sun}, \citenamefont {Cohen},\ and\ \citenamefont
  {Louie}}]{choi2002origin}%
  \BibitemOpen
  \bibfield  {author} {\bibinfo {author} {\bibfnamefont {H.~J.}\ \bibnamefont
  {Choi}}, \bibinfo {author} {\bibfnamefont {D.}~\bibnamefont {Roundy}},
  \bibinfo {author} {\bibfnamefont {H.}~\bibnamefont {Sun}}, \bibinfo {author}
  {\bibfnamefont {M.~L.}\ \bibnamefont {Cohen}}, \ and\ \bibinfo {author}
  {\bibfnamefont {S.~G.}\ \bibnamefont {Louie}},\ }\href@noop {} {\bibfield
  {journal} {\bibinfo  {journal} {Nature}\ }\textbf {\bibinfo {volume} {418}},\
  \bibinfo {pages} {758} (\bibinfo {year} {2002}{\natexlab{b}})}\BibitemShut
  {NoStop}%
\bibitem [{\citenamefont {Bohnen}\ \emph {et~al.}(2001)\citenamefont {Bohnen},
  \citenamefont {Heid},\ and\ \citenamefont {Renker}}]{bohnen2001phonon}%
  \BibitemOpen
  \bibfield  {author} {\bibinfo {author} {\bibfnamefont {K.-P.}\ \bibnamefont
  {Bohnen}}, \bibinfo {author} {\bibfnamefont {R.}~\bibnamefont {Heid}}, \ and\
  \bibinfo {author} {\bibfnamefont {B.}~\bibnamefont {Renker}},\ }\href@noop {}
  {\bibfield  {journal} {\bibinfo  {journal} {Physical Review Letters}\
  }\textbf {\bibinfo {volume} {86}},\ \bibinfo {pages} {5771} (\bibinfo {year}
  {2001})}\BibitemShut {NoStop}%
\bibitem [{\citenamefont {Morshedloo}\ \emph {et~al.}(2015)\citenamefont
  {Morshedloo}, \citenamefont {Roknabadi},\ and\ \citenamefont
  {Behdani}}]{morshedloo2015first}%
  \BibitemOpen
  \bibfield  {author} {\bibinfo {author} {\bibfnamefont {T.}~\bibnamefont
  {Morshedloo}}, \bibinfo {author} {\bibfnamefont {M.}~\bibnamefont
  {Roknabadi}}, \ and\ \bibinfo {author} {\bibfnamefont {M.}~\bibnamefont
  {Behdani}},\ }\href@noop {} {\bibfield  {journal} {\bibinfo  {journal}
  {Physica C: Superconductivity and its Applications}\ }\textbf {\bibinfo
  {volume} {509}},\ \bibinfo {pages} {1} (\bibinfo {year} {2015})}\BibitemShut
  {NoStop}%
\bibitem [{\citenamefont {Togo}\ and\ \citenamefont
  {Tanaka}(2015)}]{togo2015first}%
  \BibitemOpen
  \bibfield  {author} {\bibinfo {author} {\bibfnamefont {A.}~\bibnamefont
  {Togo}}\ and\ \bibinfo {author} {\bibfnamefont {I.}~\bibnamefont {Tanaka}},\
  }\href@noop {} {\bibfield  {journal} {\bibinfo  {journal} {Scripta
  Materialia}\ }\textbf {\bibinfo {volume} {108}},\ \bibinfo {pages} {1}
  (\bibinfo {year} {2015})}\BibitemShut {NoStop}%
\bibitem [{\citenamefont {Decker}\ and\ \citenamefont
  {Kasper}(1959)}]{decker1959crystal}%
  \BibitemOpen
  \bibfield  {author} {\bibinfo {author} {\bibfnamefont {B.}~\bibnamefont
  {Decker}}\ and\ \bibinfo {author} {\bibfnamefont {J.}~\bibnamefont
  {Kasper}},\ }\href@noop {} {\bibfield  {journal} {\bibinfo  {journal} {Acta
  Crystallographica}\ }\textbf {\bibinfo {volume} {12}},\ \bibinfo {pages}
  {503} (\bibinfo {year} {1959})}\BibitemShut {NoStop}%
\end{thebibliography}%

\end{document}